\documentclass{aa}
\usepackage{graphics}
\def\ltsima{$\; \buildrel < \over \sim \;$}
\def\simlt{\lower.5ex\hbox{\ltsima}}    
\def\gtsima{$\; \buildrel > \over \sim \;$}
\def\simgt{\lower.5ex\hbox{\gtsima}}    

\def\ref{\par\noindent\hangindent 20 pt}

\def\mincir{\ \raise -2.truept\hbox{\rlap{\hbox{$\sim$}}\raise5.truept 
\hbox{$<$}\ }}  %
\def\magcir{\ \raise -2.truept\hbox{\rlap{\hbox{$\sim$}}\raise5.truept %

\hbox{$>$}\ }}
\def\ea {et al. }
\def\cosmo{H$_0$ = 50~km~s$^{-1}$~kpc$^{-1}$ and q$_0 = 0$} 

\def\approxlt{\mathrel{\hbox{ \lower .5ex \hbox {$\sim$}
	\llap{\raise .15 ex \hbox{$<$}} }}}
\def\approxgt{\mathrel{\hbox{ \lower .5ex \hbox {$\sim$}
	\llap{\raise .15 ex \hbox{$>$}} }}}

\def\multleft#1{\hbox to size{\vbox {\halign {\lft{##}\cr #1}}\hfill}\par}
\def\multright#1{\hbox to size{\vbox {\halign {\rt{##}\cr #1}}\hfill}\par}

\def\today{\ifcase\month\or January\or February\or March\or April\or May\or
      June\or July\or August\or September\or October\or November\or December\fi
      \space\number\day, \number\year}
\def\<{\thinspace}

\def\boxit#1{\vbox{\hrule\hbox{\vrule\kern3pt\vbox{\kern3pt
	  #1 \kern3pt}\kern3pt\vrule}\hrule}}

\def\H2{\hbox{H$_{2}$~}}

\begin{document}
 
\input psfig.sty
\thesaurus{03(11.01.2; 11.02.1; 11.09.2; 11.14.1; 11.16.1; 11.19.6)}

\title{Optical imaging of the host galaxies of X--ray selected BL~Lacertae 
objects\thanks{Based on observations made with the Nordic Optical Telescope, 
operated on the island of La Palma, jointly by Denmark, Finland, Iceland, 
Norway and Sweden, in the Spanish Observatorio del Roque de los Muchachos of 
the Instituto de Astrofisica de Canarias.}}

\author{Renato Falomo\inst{1} \and Jari K. Kotilainen\inst{2} }

\institute{Osservatorio Astronomico di Padova, Vicolo dell'Osservatorio 5, 
35122 Padova, Italy; e-mail: falomo@pd.astro.it
\and Tuorla Observatory, University of Turku, V\"{a}is\"{a}l\"{a}ntie 20, 
FIN--21500 Piikki\"{o}, Finland; e-mail: jkotilai@stardust.astro.utu.fi}

\offprints{R. Falomo}

\date{Received / Accepted}
\titlerunning{Host galaxies of X-ray BL Lacs}
\authorrunning{Falomo \& Kotilainen}
\maketitle

\begin{abstract}

We investigate the properties of the host galaxies of X--ray selected (high 
frequency peaked) BL Lac objects using a large and homogeneous data set of 
high spatial resolution $R$--band observations of 52 BL Lacs in the EMSS and 
Slew samples. The redshift distribution of the BL Lacs ranges from z = 0.04 
to z$>$0.7, with average and median redshifts z = 0.26 and z = 0.24, 
respectively. Eight objects are at unknown redshift. 

We are able to resolve 45 objects out of the 52 BL Lacs. For all the well 
resolved sources, we find the host to be a luminous elliptical galaxy. In 
a few  cases a disk is not ruled out but an elliptical model is 
still preferred. 

The average absolute magnitude of the host galaxies is $<M_R(host)>$ = 
--23.9$\pm$0.6, while the average scale length of the host is $<$R(e)$>$ = 
9$\pm$5 kpc. There is no difference in the host properties between the EMSS 
and Slew samples. We find a good agreement between the results derived by the 
surveys of Wurtz et al. (ground-based data) and Urry et al. (HST data), and 
by our new deeper imaging. The average luminosity of the BL Lac hosts is 
between those of F-R I and F-R II radio galaxies in Govoni et al., supporting 
the idea that both radio galaxy types could contribute to the parent 
population. The BL Lac hosts follow the F-P relation for giant ellipticals 
and exhibit a modest luminosity evolution with redshift. Finally, we find a 
slight correlation between the nuclear and host luminosity and a bimodal 
distribution in the nuclear/host luminosity ratio. 


\end{abstract}

\section{Introduction}

BL Lacertae objects are the most extreme class of active galactic nuclei 
(AGN), exhibiting strong, rapidly variable polarization and continuum 
emission, and core-dominated radio emission with apparent superluminal motion 
(see e.g. Kollgaard et al. 1992 for references). These properties have led to 
the commonly accepted view that BL Lacs are dominated by Doppler-boosted 
synchrotron emission from a relativistic jet nearly along our line-of-sight 
(Blandford \& Rees 1978). The line emission of BL Lacs is absent or weak, 
making their redshift determination rather difficult.

In the current unified models of radio-loud AGN (e.g. Urry \& Padovani
1995), BL Lacs are identified as low luminosity, core-dominated F-R I
(Fanaroff \& Riley 1974) radio galaxies (RG) viewed nearly along the
axis of the relativistically boosted jet. This model appears supported by
the comparison of their extended radio luminosity and morphology
(e.g. Perlman \& Stocke 1993), host galaxy luminosity and morphology
(e.g. Wurtz et al. 1996; hereafter WSY96) and space density and beamed
luminosity functions (e.g. Padovani \& Urry 1990; Morris et al. 1991;
Celotti et al. 1993).

Knowledge of the properties of the host galaxies and environments where AGN 
live is fundamental for the understanding of the formation of AGN in 
galaxies. Comparison of orientation-independent properties of BL Lacs, such 
as the host galaxies and environment, with those of RGs, allow one to test 
the unified model based on orientation (Antonucci 1993; Urry \& Padovani 
1995). The frequency of close companions will determine whether interactions 
are important for triggering of the BL Lac activity, as seems to be the case 
for quasars (e.g. Heckman 1990; Hutchings \& Neff 1992). Possible 
cosmological evolution in the properties of the hosts and environments can be 
studied by comparing AGN properties at different redshifts. Finally, the 
alternative gravitational lensing hypothesis for BL Lac activity (Ostriker \& 
Vietri 1990) can be tested by measuring the predicted offsets between the BL 
Lacs and their host galaxies.

Recent studies of BL Lac hosts and close environments from the ground
and with the HST (see Pesce et al 1995; Falomo 1996; Owen et al. 1996;
WSY96; Wurtz et al. 1997; Falomo et al. 1997; Scarpa et al 1999a; Urry
et al. 1999a; Urry et al. 1999b, hereafter U99b; Heidt et al. 1999)
have shown that their host galaxies are luminous (M$_R \sim $ --23
to --24 ) and large (R(e) = 10$\pm$7 kpc) elliptical galaxies. They are on
average $\sim$1 mag brighter than L$^*$ galaxies (e.g.  Mobasher et
al. 1993) and of similar luminosity or slightly fainter than the
brightest cluster galaxies (BCG; e.g. Hoessel et
al. 1980). Intriguingly, BL Lac hosts appear slightly fainter than F-R
I hosts and resemble better typical F-R II hosts.  Detections or
claims of disk hosts and/or inner disks components have been made in a
handful of sources (e.g. MS 0205.7+3509 Stocke Wurtz and Perlman 1995;
OQ 530 Abraham et al 1991; PKS 1413+135, McHardy et al.
1991; PKS 0548-322, Falomo et al. 1995; 1ES 1959+650, Heidt et
al. 1999;). While it is not unreasonable to find small inner disk
structures similarly to what found in normal ellipticals, the objects
with a disk dominated host galaxy are rather controversial (for
details see: Falomo et al 1997; Stocke et al 1992; Lamer, Newsam, and
McHardy 1999; Scarpa et al 1999a,b).

Close companions have been found around many BL Lacs, some
with signs of interaction, but the physical association has been
confirmed only in some cases through spectroscopic measurements. BL
Lacs usually reside in poor clusters, unlike F-R I RGs. A modified
unified model has therefore been proposed in which BL Lacs either
include partly F-R IIs in addition to F-R Is, or BL Lacs avoid the
brightest F-R Is and the F-R Is in rich clusters at low redshift
(WSY96).  This scenario seems also supported by recent measurements of
radio polarization of BL Lacs (Stanghellini et al 1997) that show
signature of a F-R II population.

Similarly to the case of quasars (e.g. Hutchings et al. 1994;
R\"onnback et al. 1996; Bahcall et al. 1997), the study of BL Lac host
galaxies is rather problematic because of the presence of the bright
nucleus that swamps the light of the host galaxy, although the
nucleus/host luminosity ratio of the BL Lacs is smaller than that of
quasars. The use of HST images can substantially improve the ability
to study the host close to the nucleus, however, as we shall show in
Section 4.2.2, HST data are usually not deep enough to properly
investigate the external fainter regions of the host galaxies (see
also e.g. Hutchings et al. 1994; Bahcall et al 1997).

In this paper we report on a large, homogeneous data set of observations of 
BL Lacs secured mainly with sub-arcsec resolution from the Nordic Optical 
Telescope (NOT) in La Palma. We have observed the complete sample of 26 X-ray 
selected (or high frequency peaked, HBL) BL Lacs derived from the {\em 
Einstein} Medium Sensitivity Survey (EMSS; Stocke et al. 1991). This sample 
is almost complete for multiwavelength information and redshift (although 
some redshifts are tentative and probably wrong, see the Appendix) and most of 
the objects are at z$<$0.3, ensuring the detection of the host in most cases 
with reasonable observing time. Additionally, we obtained images for 26 BL 
Lacs in the {\em Einstein} Slew Survey (Perlman et al 1996). These BL Lacs 
are also of HBL type and are compared with the EMSS targets. Based on the 
current data, we have previously published a separate study of the peculiar 
BL Lac object MS 0205.7+3509 (Falomo et al. 1997).

The structure of this paper is as follows: In Section 2 we describe the 
samples of the objects observed. In Section 3 we give details of the 
observations, data reduction and describe the image analysis. Results and 
discussion on the overall properties of the host galaxies, including 
comparison with previous studies of BL Lacs and other types of AGN are given 
in Section 4. The main conclusions from this study are summarized in 
Section 5. In the Appendix, we give comments for individual objects. 
Throughout this paper, \cosmo ~are adopted.

\section{The samples}

\subsection{The EMSS sample}

The EMSS survey (Gioia et al. 1990; Morris et al 1991; Stocke et al. 1991; 
Maccacaro et al. 1994) is a flux--limited complete sample of faint X--ray 
sources discovered serendipitously in numerous {\em Einstein} Imaging 
Proportional Counter (IPC) fields centered on high galactic latitude (b $>$ 
20 $\deg$) targets. It covers 780 $\deg^2$ in the 0.3 -- 3.5 keV soft X--ray 
band with limiting sensitivity ranging from $\sim$5 $\times$ 10$^{-14}$ to 
$\sim$3 $\times$ 10$^{-12}$ erg cm$^{-2}$ s$^{-1}$. The EMSS includes 34 BL 
Lac objects and 4 BL Lac candidates (as listed in Padovani \& Giommi 1995), 
selected to have the observed equivalent width of any emission line 
EW $<$5 $\AA$. 
Moreover, if a Ca II H+K break is present due to starlight in the BL 
Lac host galaxy, its contrast must be $\leq$ 25 $\%$, much less than for a 
typical giant non-active elliptical galaxy ($\sim$50 \%).

\subsection{The Slew survey sample}

The IPC Slew survey (Perlman et al. 1996) was constructed using the {\em 
Einstein} slew data taken when the satellite was moving from one target to 
the next (Elvis et al. 1992) and covers a large fraction of the sky with 
limiting sensitivity ranging from $\sim$5 $\times$ 10$^{-12}$ to $\leq$1 
$\times$ 10$^{-12}$ erg cm$^{-2}$ s$^{-1}$ in the 0.3 -- 3.5 keV soft X--ray 
band. Padovani \& Giommi (1995) list a total of 60 Slew BL Lacs and 9 BL Lac 
candidates extracted from the Slew survey adopting the same classification 
criteria as for the EMSS survey. 


\subsection{Our selection criteria}

The observed BL Lacs were selected from the EMSS and Slew samples to have 
declination $\delta \geq$--15 $\deg$, to be observable at the NOT. This limit 
excluded five BL Lacs and three BL Lac candidates of the EMSS sample. Of the 
remaining 29 BL Lacs and one BL Lac candidate, 28 (93 \%) were observed at 
the NOT, the only exceptions being MS 1019.0+5139 and MS 1207.9+3945.

In the Slew sample, seven BL Lacs and two BL Lac candidates do not satisfy 
our declination limit. Furthermore, of the remaining 53 BL Lacs and 7 
candidates, 14 are of low frequency peak (LBL) type while 4 belong also to 
the EMSS sample. Of the final sample of 35 Slew BL Lacs and 7 Slew BL Lac 
candidates, 26 (62 \%) were observed at the NOT. The selection of the Slew 
objects observed was based only on observability conditions. General 
properties of the observed BL Lacs are given in Table 1, columns (1)-(7), 
where column (1) gives the name of the object, column (2) the redshift, 
column (3) the apparent $V$--band magnitude, columns (4) and (5) the 5 GHz 
and 2 keV flux densities, respectively, and columns (6) and (7) the 
optical--X-ray and radio--optical spectral indices, respectively.


\begin{table*}
\begin{center}
\caption{General properties of the BL Lacs and journal of the observations.}
\begin{tabular}{llllllllllll}
\hline\\
Name & z & V & S(5 GHz) & S(2 keV) & $\alpha$(O--X) & $\alpha$(R--O) & Date & 
T(int) & Sky  & Seeing & A(r)\\
     &   &   & (mJy)   & ($\mu$Jy) &               &                &       & 
(sec)& (mag) & ($\prime\prime$) & (mag)\\
1ES 0033+595   & ...   & 19.5 &  ~66.0 & 2.22 & 0.45 & 0.61 & 21/09/98 &  ~600 & 21.0  & 0.64 & 1.94 \\ 
1ES 0120+340   & 0.272 & 15.2 &  ~33.6 & 1.86 & 1.06 & 0.28 & 23/12/95 & 1200 & 19.72 & 0.51 & 0.24 \\ 
MS 0122.1+0903 & 0.339 & 20.0 &   ~~1.4 & 0.15 & 0.91 & 0.34 & 22/09/98 & 1800 & 21.01 & 0.58 & 0.20 \\
MS 0158.5+0019 & 0.229: & 18.0 &  ~11.3 & 0.67 & 0.88 & 0.36 & 22/09/98 & 3600 & 20.9  & 0.63 & 0.12 \\
MS 0205.7+3509 & 0.318: & 19.2 &   ~~3.6 & 0.11 & 1.08 & 0.36 & 23/12/95 & 2400 & 19.97 & 0.68 & 0.29 \\
1ES 0229+200   & 0.139 & 14.7 &  ~49.1 & 1.13 & 1.22 & 0.27 & 22/09/98 & 2400 & 21.01 & 0.60 & 0.41 \\
MS 0257.9+3429 & 0.247 & 18.5 &  ~10.0 & 0.25 & 1.02 & 0.40 & 23/12/95 & 1800 & 20.95 & 0.53 & 0.41 \\
MS 0317.0+1834 & 0.190 & 18.1 &  ~17.0 & 2.56 & 0.71 & 0.41 & 24/12/95 & 1800 & 20.88 & 0.62 & 0.46 \\
1ES 0347-121   & 0.188 & 18.2 &   ~~9.0 & 2.32 & 0.64 & 0.36 & 23/09/98 & 1800 & 21.01 & 1.12 & 0.17 \\
1ES 0414+009   & 0.287 & 17.5 &  ~70.0 & 3.64 & 0.67 & 0.47 & 24/02/98 & 2400 & 20.2  & 0.87 & 0.41 \\ 
MS 0419.3+1943 & 0.512: & 20.3 &   ~~8.0 & 0.53 & 0.62 & 0.50 & 24/12/95 & 3600 & 19.9  & 0.73 & 0.84 \\
1ES 0446+449   & 0.203: & 18.5 & 139.4 & 0.63 & 0.81 & 0.60 & 24/12/95 & 1800 & 20.79 & 0.59 & 2.52 \\
1ES 0502+675   & 0.341 & 17.0 &  ~32.7 & 1.36 & 0.92 & 0.37 & 21/09/98 & 3600 & 21.01 & 0.78 & 0.53 \\
1ES 0525+713   & 0.249: & 19.0 &  ~~9.0 & 0.80 & 0.69 & 0.42 & 21/09/98 & 1200 & 20.9  & 0.68 & 0.41 \\ 
MS 0607.9+7108 & 0.267 & 19.6 &  ~18.2 & 0.27 & 0.86 & 0.52 & 24/12/95 & 3300 & 20.67 & 0.67 & 0.41 \\
1ES 0647+250   & ...   & 15.8 &  ~73.4 & 2.36 & 1.01 & 0.35 & 21/09/98 &  ~900 & 20.9  & 0.65 & 0.65 \\
MS 0737.9+7441 & 0.315 & 16.9 &  ~24.0 & 0.46 & 1.10 & 0.40 & 24/12/95 & 1500 & 20.68 & 0.77 & 0.17 \\
1ES 0806+524   & 0.136: & 15.0 & 171.9 & 1.38 & 1.22 & 0.36 & 24/02/98 & 3000 & 20.73 & 0.96 & 0.20 \\
MS 0922.9+7459 & 0.638: & 19.7 &   ~~3.3 & 0.22 & 0.90 & 0.39 & 24/12/95 & 3600 & 20.66 & 0.71 & 0.12 \\
1ES 0927+500   & 0.188 & 17.2 &  ~18.3 & 0.67 & 1.00 & 0.34 & 24/12/95 & 2100 & 20.84 & 0.93 & 0.08 \\
MS 0950.9+4929 & $>$0.5 & 19.3 &   ~~3.3 & 0.21 & 0.88 & 0.36 & 02/06/95 & 1320 & 20.26 & 0.83 & 0.05 \\
MS 0958.9+2102 & 0.334 & 19.8 &   ~~1.5 & 0.04 & 1.16 & 0.34 & 02/06/95 & 1320 & 19.88 & 0.88 & 0.12 \\
1ES 1011+496   & 0.210 & 16.1 & 286.0 & 0.54 & 1.21 & 0.48 & 24/02/98 & 2400 & 20.75 & 1.66 & 0.05 \\
1ES 1028+511   & 0.239: & 16.6 &  ~44.2 & 1.88 & 0.92 & 0.37 & 24/12/95 & 1500 & 20.96 & 0.57 & 0.05 \\
1ES 1106+244   & ...   & 18.7 &  ~18.1 & 0.56 & 0.80 & 0.45 & 25/02/98 & 2400 & 20.75 & 2.02 & 0.05 \\
1ES 1118+424   & 0.124: & 17.0 &  ~35.0 & 1.41 & 0.91 & 0.38 & 24/12/95 & 1200 & 21.04 & 0.60 & 0.10 \\
1ES 1212+078   & 0.130 & 16.0 &  ~94.0 & 0.27 & 1.25 & 0.43 & 25/02/98 & 2400 & 20.23 & 1.53 & 0.08 \\
1ES 1218+304   & 0.182: & 16.4 &  ~56.0 & 2.51 & 0.90 & 0.37 & 25/02/98 & 2400 & 20.39 & 0.98 & 0.08 \\
MS 1221.8+2452 & 0.218: & 17.6 &  ~26.4 & 0.26 & 1.20 & 0.41 & 31/05/95 & 1200 & 19.95 & 1.14 & 0.10 \\
MS 1229.2+6430 & 0.164 & 16.9 &  ~42.0 & 0.70 & 1.15 & 0.39 & 01/06/95 & 1200 & 20.83 & 0.59 & 0.10 \\
MS 1235.4+6315 & 0.297 & 18.6 &   ~~7.0 & 0.39 & 0.99 & 0.37 & 01/06/95 & 1200 & 20.78 & 0.70 & 0.08 \\
1ES 1255+244   & 0.141 & 15.4 &   ~~7.4 & 3.66 & 1.30 & 0.17 & 24/12/95 &  ~900 & 20.36 & 0.58 & 0.08 \\
MS 1256.3+0151 & ...   & 20.0 &   ~~8.0 & 0.05 & 1.11 & 0.49 & 31/05/95 & 1200 & 19.41 & 1.34 & 0.08 \\
MS 1402.3+0416 & 0.344: & 17.1 &  ~20.8 & 0.10 & 1.33 & 0.34 & 03/06/95 & 1920 & 20.6  & 0.84 & 0.10 \\
MS 1407.9+5954 & 0.495 & 19.7 &  ~16.5 & 0.41 & 0.81 & 0.52 & 31/05/95 & 1800 & 19.69 & 1.36 & 0.08 \\
MS 1443.5+6349 & 0.299 & 19.6 &  ~11.6 & 0.33 & 0.85 & 0.49 & 02/06/95 & 1920 & 20.82 & 0.76 & 0.08 \\
MS 1458.8+2249 & 0.235: & 16.8 &  ~29.8 & 0.22 & 1.35 & 0.36 & 02/06/95 & 1320 & 20.97 & 0.64 & 0.17 \\
1ES 1517+656   & $>$0.7 & 15.9 &  ~39.0 & 1.19 & 1.11 & 0.30 & 21/09/98 & 2400 & 19.5  & 0.70 & 0.10 \\
MS 1534.2+0148 & 0.312 & 18.7 &  ~34.0 & 0.43 & 0.94 & 0.51 & 01/06/95 & 1920 & 20.41 & 0.65 & 0.24 \\
MS 1552.1+2020 & 0.222 & 17.7 &  ~37.5 & 0.89 & 0.97 & 0.44 & 02/06/95 & 2100 & 20.88 & 0.65 & 0.17 \\
MS 1704.9+6046 & 0.280 & 19.1 &   ~~1.8 & 0.10 & 1.13 & 0.30 & 31/05/95 & 1800 & 19.31 & 1.95 & 0.10 \\
MS 1757.7+7034 & 0.407 & 18.3 &   ~~7.2 & 0.48 & 0.98 & 0.35 & 24/09/98 & 2400 & 20.9  & 1.20 & 0.20 \\
1ES 1853+671   & 0.212 & 16.4 &  ~12.1 & 0.28 & 1.19 & 0.28 & 21/09/98 & 2400 & 21.01 & 0.64 & 0.26 \\
1ES 1959+650   & 0.048: & 13.7 & 251.6 & 3.64 & 1.19 & 0.32 & 23/09/98 & 1800 & 20.99 & 0.62 & 0.48 \\
1ES 2037+521   & ...   & 19.0 &  ~32.5 &      &      &      & 22/09/98 & 2400 & 20.91 & 0.77 & 2.21 \\ 
MS 2143.4+0704 & 0.237 & 18.0 &  ~50.0 & 0.46 & 1.03 & 0.49 & 01/06/95 & 1320 & 20.57 & 0.82 & 0.22 \\
1ES 2321+419   & 0.059: & 17.0 &  ~19.0 & 0.27 & 1.19 & 0.32 & 24/09/98 & 1200 & 21.14 & 0.69 & 0.44 \\
1ES 2326+174   & 0.213 & 16.8 &  ~18.4 & 0.56 & 1.02 & 0.35 & 21/09/98 & 1800 & 20.9  & 0.70 & 0.17 \\
MS 2336.5+0517 & ...   & 20.3 &   ~~4.9 & 0.10 & 0.93 & 0.47 & 24/12/95 & 3600 & 18.84 & 0.61 & 0.26 \\
1ES 2343-151   & 0.226 & 19.2 &   ~~8.2 & 0.30 & 0.83 & 0.42 & 23/09/98 & 3000 & 20.9  & 0.85 & 0.10 \\
1ES 2344+514   & 0.044 & 15.5 & 215.2 & 1.14 & 1.18 & 0.41 & 22/09/98 & 1200 & 20.9  & 0.99 & 0.74 \\
MS 2347.4+1924 & 0.515 & 20.8 &   ~~3.2 & 0.10 & 0.86 & 0.47 & 23/09/98 & 1800 & 20.9  & 0.65 & 0.20 \\
\hline\\
\end{tabular}
\end{center}
\end{table*}

The redshift distribution of the observed objects from EMSS and Slew is shown 
in Fig. 1. The average redshifts of the BL Lacs with known redshift in the 
samples are: 0.319$\pm$0.133 (EMSS, all); 0.314$\pm$0.120 (EMSS, observed); 
0.201$\pm$0.124 (Slew, all) and 0.190$\pm$0.091 (Slew, observed). It can be 
seen that a) the observed and full samples do not differ significantly in 
their redshift distribution and b) the Slew survey tends to select BL Lacs at 
somewhat lower redshift than the EMSS survey because of the brighter X-ray 
flux limit of the Slew survey.

\begin{figure}
\psfig{file=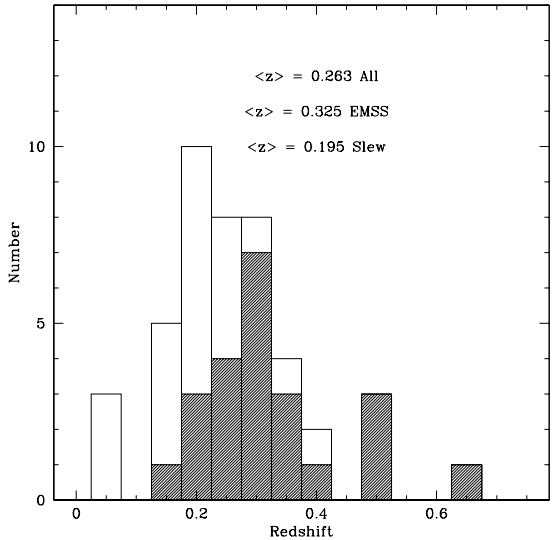,width=9cm,height=9cm}
\caption{Redshift distribution of the observed BL Lac objects. Hatched
area show EMSS objects}
\end{figure}

\section{Observations, data reduction and analysis}

Optical images were obtained during four observing runs using the 2.5m NOT 
telescope at La Palma. We used the BroCam camera (1024$^2$ px, 0\farcs176 
px$^{-1}$) for observations in 1995 while the HiRAC camera (2048$^2$ px, 
0.\farcs11 px$^{-1}$) was used for observations in 1998 (for details, see 
Table 2). In all observations the Cousins $R$ filter was used to image the 
objects. The observations were performed mostly during photometric conditions 
and photometric calibration of each night was obtained from frequent 
observations of Landolt (1992) standard stars. Some objects were imaged 
during non photometric conditions, therefore we secured additional short 
exposure images of these targets during photometric nights to calibrate these 
frames using reference stars. Seeing conditions were generally very good with 
average and median seeing FWHM = 0.\farcs84 and 0.\farcs70, respectively.

\begin{table}
\begin{center}
\caption{Description of the observing runs.}
\begin{tabular}{llll}
\hline
Date & Instrument/CCD & $\prime\prime$/px & Photometry \\
\hline
31/5-3/6/95 & BroCam/Tk1024A & 0.176 &          good\\
20-23/12/95 & BroCam/Tk1024A & 0.176 &          No (20-22/12)\\
            &                &       &          Yes (23/12)\\
24-26/2/98  & HiRAC/Loral & 0.11 &             poor\\
21-24/9/98  & HiRAC/Loral & 0.11 &             good\\
\hline\\
\end{tabular}
\end{center}
\end{table}
  
In most cases both a short (typically 2 minutes) and a long (ranging from 10 
to 60 minutes, average 20 minutes) integration were obtained. The short and 
long exposures were then combined to form an image of the target well exposed 
in the external (fainter) regions while avoiding saturation of the bright 
nucleus. In some cases where the nucleus was much brighter than the 
surrounding nebulosity, several intermediate length integrations were 
combined in order to obtain a final well exposed image. Moreover, the 
combination of multiple images allowed for identifying and removing cosmic 
ray hits.

In Table 1, columns (8)-(12), we give a journal of the observations with 
details for each object. Column (8) gives the date of observation, columns 
(9) and (10) report the total exposure time and the sky brightness, while in 
columns (11) and (12) the seeing FWHM measured from stellar images and the 
Galactic extinction used are given.

Data reduction was performed following standard procedures (including bias 
subtraction, flat fielding and cosmic ray rejection) available in 
IRAF{\footnote{IRAF is distributed by the National Optical Astronomy 
Observatories, which are operated by the Association of Universities for 
Research in Astronomy, Inc., under cooperative agreement with the National 
Science Foundation}}. The level of the sky was derived sampling several 
regions over the image and checking for residual gradients in the background 
level. Since no  significant gradient was found  a single flux 
level for the sky was used. From these final images we have extracted for 
each object the azimuthally averaged radial brightness profile down to a 
surface brightness $\mu_R$ $\sim$ 26 mag arcsec$^{-2}$. Any obvious extra 
features (e.g. companions and foreground stars) were removed (masked) from 
the image in order to avoid contamination of the radial profile. In order to 
derive the shape of the point spread function (PSF), we have similarly 
extracted the radial profile of a large number of stars in each field. Since 
the field of view of our images is sufficiently large, this was always easily 
obtained using the object frame. Many stellar profiles were combined in order 
to obtain a good PSF both in the core and in the wings.

Subsequent analysis consisted in comparing each profile of the BL Lac object 
with its PSF and, for the resolved sources, fitting the observed luminosity 
profile with a model. We used the simple approach of assuming the observed 
object is composed of a nuclear (unresolved) source, described by a PSF, plus 
a galaxy, modeled by either a de Vaucouleurs ($\mu$(r) $\propto$ r$^{0.25}$) 
or an exponential disk ($\mu$(r) $\propto$~~r), convolved with the proper 
PSF. More complex models, for example a generalized de Vaucouleurs law (e.g. 
the Sersic law $\mu \propto r^{1/n}$), increases the number of free 
parameters and do not offer real advantages for the characterization of the 
host galaxies.

From the best fit of the profiles we have determined the parameters of
the host galaxy ($\mu_o$, r$_e$ , total magnitude) as well as the
magnitude of the nuclear source. These parameters are reported in
Table 3, columns (1)-(7), where column (1) gives the name of the
object, column (2) its redshift, columns (3) and (4) the central
surface brightness of bulge component and its scale length, columns
(5)--(6) the apparent magnitude of the nuclear and galactic
components, and column (7) the reduced $\chi^2$ of the best
fit. Absolute quantities were derived after applying correction for
Galactic extinction and redshift (K-correction). The former was
determined using the Bell Lab Survey of neutral hydrogen N$_H$
converted to E$_B-V$ (Stark et al. 1992; Shull \& Van Steenberg 1985),
while the latter was computed from the model of Coleman et al
(1980) for elliptical galaxies.

\begin{table*}
\begin{center}
\caption{Best-fit parameters of the profile fits and properties of the host galaxies.}
\begin{tabular}{lllllllllllll}
\hline
Name       & z & $\mu_0$ & r$_e$ & m$_{nuc}$ & m$_{gal}$ &  $\chi^2$ &
 K--cr. & M$_{PSF}$ & M(host) & R(e) & $\mu$(E) & Note\\ 
 & &  & ($\prime\prime$) & (mag) & (mag) &    & & (mag)  & (mag) &
 (kpc) &   & \\
\hline
1ES 0033+595   & ...    & 12.00 & ~0.40 & 18.88 &  18.92  & ~1.50 &...      & ...    &  ...   &  ...  & ...   & \\ 
1ES 0120+340   & 0.272  & 14.97 & ~3.40 & 16.39 &  17.24  & ~7.17 &
0.29     & -25.19 & -24.63 & 18.89 & 21.72& a) \\
MS 0122.1+0903 & 0.339  & 13.60 & ~0.90 & 15.43 &  18.75  & ~0.52 & 0.38     & -26.65 & -23.70 & 5.79  & 20.08&  \\
MS 0158.5+0019 & 0.229: & 14.70 & ~2.05 & 18.82 &  18.07  & ~0.26 & 0.32     & -22.87 & -23.94 & 12.15 & 21.45&  \\
MS 0205.7+3509 & 0.318: & 14.50 & ~1.20 & 17.53 &  19.03  & ~0.06 &
0.35     & -24.48 & -23.33 & 7.41  & 20.99& a)  \\
1ES 0229+200   & 0.139  & 13.84 & ~4.00 & 18.25 &  15.76  & ~0.04 & 0.13     & -21.91 & -24.53 & 13.33 & 21.06&  \\
MS 0257.9+3429 & 0.247  & 14.44 & ~2.05 & 19.07 &  17.81  & ~1.32 & 0.27     & -22.45 & -23.98 & 10.64 & 21.13&  \\
MS 0317.0+1834 & 0.190  & 12.70 & ~1.10 & 19.01 &  17.42  & ~4.10 & 0.20     & -21.93 & -23.72 & 4.70  & 19.61& a) \\
1ES 0347-121   & 0.188  & 11.80 & ~0.65 & 19.37 &  17.66  & ~1.16 & 0.20     & -21.26 & -23.17 & 2.76  & 19.01&  \\
1ES 0414+009   & 0.287  & 12.54 & ~1.00 & 16.97 &  17.47  & ~0.41 & 0.31     & -24.91 & -24.72 & 5.76  & 19.05&  \\
MS 0419.3+1943 & 0.512: & 14.10 & ~0.74 & 18.65 &  19.68  & ~3.31 & 0.75     & -25.12 & -24.84 & 6.05  & 19.04&  \\
1ES 0446+449   & 0.203: & 16.11 & 22.90 & ...  &  14.24  & ...  & 0.21  & ...    & -29.13 & 102.92 & 20.90&  \\
1ES 0502+675   & 0.341  & 10.98 & ~0.35 & 16.83 &  18.19  & ~5.25 & 0.38     & -25.60 & -24.62 & 2.26  & 17.12& a) \\
1ES 0525+713   & 0.249: & 14.35 & ~2.30 & ...   &  17.47  & ~1.27 & 0.27     & ...    & -24.34 & 12.01 & 21.03&  \\
MS 0607.9+7108 & 0.267  & 14.1  & ~1.85 & 18.31 &  17.69  & ~1.35 & 0.29    & -23.40 & -24.31 & 10.1 & 20.7&  \\
1ES 0647+250   & ...    & ...  & ...  & 15.03 &   ...   & 12.95 & 0.21    & -26.26 & ...    & ...   & ...  &  \\
MS 0737.9+7441 & 0.315  & 12.97 & ~1.10 & 18.29 &  17.69  & ~4.01 & 0.34     & -23.58 & -24.52 & 6.75 & 19.60&  \\
1ES 0806+524   & 0.136: & 12.05 & ~1.70 & 15.77 &  15.83  & ~2.32 & 0.13     & -24.16 & -24.24 & 5.63 & 19.48&  \\
MS 0922.9+7459 & 0.638: & 15.26 & ~1.03 & 20.34 &  20.12 & 16.87 & 1.15     & -23.30 & -24.67 & 9.40 & 20.17 & \\
1ES 0927+500   & 0.188  & 15.50 & ~3.00 & 17.65 &  18.04  & ~2.20 & 0.20     & -22.89 & -22.70 & 12.72 & 22.80&  \\
MS 0950.9+4929 & $>$0.5 & ...   &  ... & 18.78 &    ...  & ~0.04 & 0.22     & -21.95 & ...   & ...   & ... &   \\
MS 0958.9+2102 & 0.334  & 15.18 & ~1.82 & 20.66 &  18.81  & ~0.10 & 0.37     & -21.30 & -23.52 & 11.60 & 21.76&  \\
1ES 1011+496   & 0.210  & 14.37 & ~2.73 & 16.07 &  17.12 & 15.92 & 0.21     & -24.58 & -23.74 & 12.13 & 21.64&  \\
1ES 1028+511   & 0.239: & 13.75 & ~1.30 & 16.69 &  18.11  & ~5.27 & 0.26     & -24.39 & -23.23 & 6.59 & 20.83&  \\
1ES 1106+244   & ...    & 15.60 & ~1.40 & 18.71 &   19.8  & ~5.35 &
0.48     & -23.64 & -23.03 & 9.97 & 21.93&  a) \\
1ES 1118+424   & 0.124: & ...   &      & 16.88 &   ...  & ~1.83 & 0.11     & -22.71 & ...   & ...  & ... &   \\
1ES 1212+078   & 0.130  & 14.64 & ~5.51 & 17.26 &  15.86  & ~0.33 & 0.12     & -22.42 & -23.94 & 17.38 & 22.23&  \\
1ES 1218+304   & 0.182: & 13.55 & ~2.17 & 16.32 &   16.8  & ~0.18 & 0.19     & -24.14 & -23.85 & 8.97 & 20.88&  \\
MS 1221.8+2452 & 0.218: & 12.46 & ~0.62 & 17.76 &  18.43 & 75.64 & 0.23     & -23.15 & -22.71 & 2.94 & 19.60&  \\
MS 1229.2+6430 & 0.164  & 13.51 & ~2.90 & 17.35 &  16.13  & ~2.06 & 0.17     & -22.89 & -24.28 & 11.05 & 20.91&  \\
MS 1235.4+6315 & 0.297  & 13.80 & ~1.39 & 19.20 &  18.01  & ~0.11 & 0.32     & -22.43 & -23.95 & 8.20 & 20.60&  \\
1ES 1255+244   & 0.141  & 13.10 & ~1.90 & 18.27 &  16.64  & ~1.23 & 0.14     & -21.59 & -23.36 & 6.41 & 20.63&  \\
MS 1256.3+0151 & ...    & ...   & ...  & 19.50 &   ...  & ~0.14 & ...      & ...    & ...     & ...  & ...  &  \\
MS 1402.3+0416 & 0.344: & ...   & ...  & 16.88 &   ...  & ~0.60 & 0.39     & -25.14 & ...   & ...   & ...  &  \\
MS 1407.9+5954 & 0.495  & 14.80 & ~1.00 & 18.69 &  19.73  & ~5.53 & 0.70     & -24.23 & -23.89 &  8.04 & 20.60&  \\
MS 1443.5+6349 & 0.299  & 15.49 & ~2.90 & 19.38 &  18.11  & ~2.90 & 0.32     & -22.27 & -23.86 & 17.18 & 22.28&  \\
MS 1458.8+2249 & 0.235: & 10.50 & ~0.46 & 16.27 &  17.12  & ~0.98 & 0.25     & -24.89 & -24.29 & 2.30 & 17.49&  \\
1ES 1517+656   & $>$0.7 & ...  & ...  & 16.38 &    ...  & ~6.86 & 1.35     & -27.49 & ...    & ...  & ...  &  \\
MS 1534.2+0148 & 0.312  & 15.81 & ~4.05 & 19.22 &   17.7 & 10.38 &
0.34     & -22.70 & -24.56 & 24.68 & 22.38&  b) \\
MS 1552.1+2020 & 0.222  & 13.80 & ~2.35 & 17.97 &  16.87  & ~0.59 & 0.24     & -23.05 & -24.39 & 11.29 & 20.84&  \\
MS 1704.9+6046 & 0.280  & 14.95 & ~2.40 & 19.49 &  17.98  & ~0.20 & 0.30     & -22.02 & -23.83 & 13.60 & 21.80&  \\
MS 1757.7+7034 & 0.407  & 12.14 & ~0.30 & 17.89 &  19.68  & ~0.84 & 0.50     & -24.65 & -23.36 & 2.16&  18.28&  a)\\
1ES 1853+671   & 0.212  & 15.64 & ~2.50 & 19.47 &  18.58  & ~0.51 & 0.23     & -21.53 & -22.65 & 11.61 & 22.64&  \\
1ES 1959+650   & 0.048: & 13.25 & ~4.10 & 14.86 &  15.12 & 10.85 & 0.04     & -22.97 & -22.75 & 5.34 & 20.85&  \\
1ES 2037+521   & ...    & 14.86 & ~5.80 & 19.64 &  15.97  & ~2.17 & 0.04     & -20.01 & -23.72 & 7.84 & 20.72&  \\
MS 2143.4+0704 & 0.237  & 14.62 & ~2.17 & 18.24 &  17.87  & ~0.18 & 0.26     & -22.99 & -23.62 & 10.94 & 21.54&  \\
1ES 2321+419   & 0.059: & ...   & ...  & 17.31 &   ...  & ~1.35 & 0.05     & -20.94 & ...   & ...   & ...  &  \\
1ES 2326+174   & 0.213  & 13.06 & ~1.60 & 18.41 &  16.97  & ~2.42 & 0.23     & -22.51 & -24.18 & 7.46 & 20.15 & \\
MS 2336.5+0517 & ...    & 12.60 & ~0.60 & 19.83 &  18.64  & ~0.08 & ...      & ...    & ...     & ...  & ...  &  \\
1ES 2343-151   & 0.226  & 13.27 & ~1.60 & 20.36 &  17.18  & ~0.67 &
0.24     & -20.63 & -24.05 & 7.79 & 20.37&  b) \\
1ES 2344+514   & 0.044  & 12.87 & ~5.80 & 17.00 &  13.98  & ~0.25 & 0.03     & -20.89 & -23.95 & 6.96 & 20.24& \\ 
MS 2347.4+1924 & 0.515  & 13.27 & ~0.40 & 21.89 &  20.19  & ~0.75 & 0.76     & -21.26 & -23.72 & 3.28 & 18.83&  \\
\hline\\
\multicolumn{13}{l}{a) elliptical fit preferred but disk fit not ruled
out} \\
\multicolumn{13}{l}{b) poor  fit in the external outer region} \\
\end{tabular}
\end{center} 
\end{table*}

\section{Results and discussion}

\subsection{Absolute magnitude and scale length}

In Fig. 2 we report for each object the radial luminosity profile
together with its best-fit decomposition into nucleus and host galaxy
components. We are able to resolve 45 objects out of the 52 observed
sources. For almost all of the clearly resolved sources we are able to
fit the luminosity profile of the host galaxy with an elliptical model
while a disk model gave a significantly worse fit. For a few distant
and marginally resolved sources (see Table 3 and Fig. 2), we cannot
rule out a disk model but even in these cases an elliptical model is a
good representation of the host galaxy.

\begin{figure*}
\psfig{file=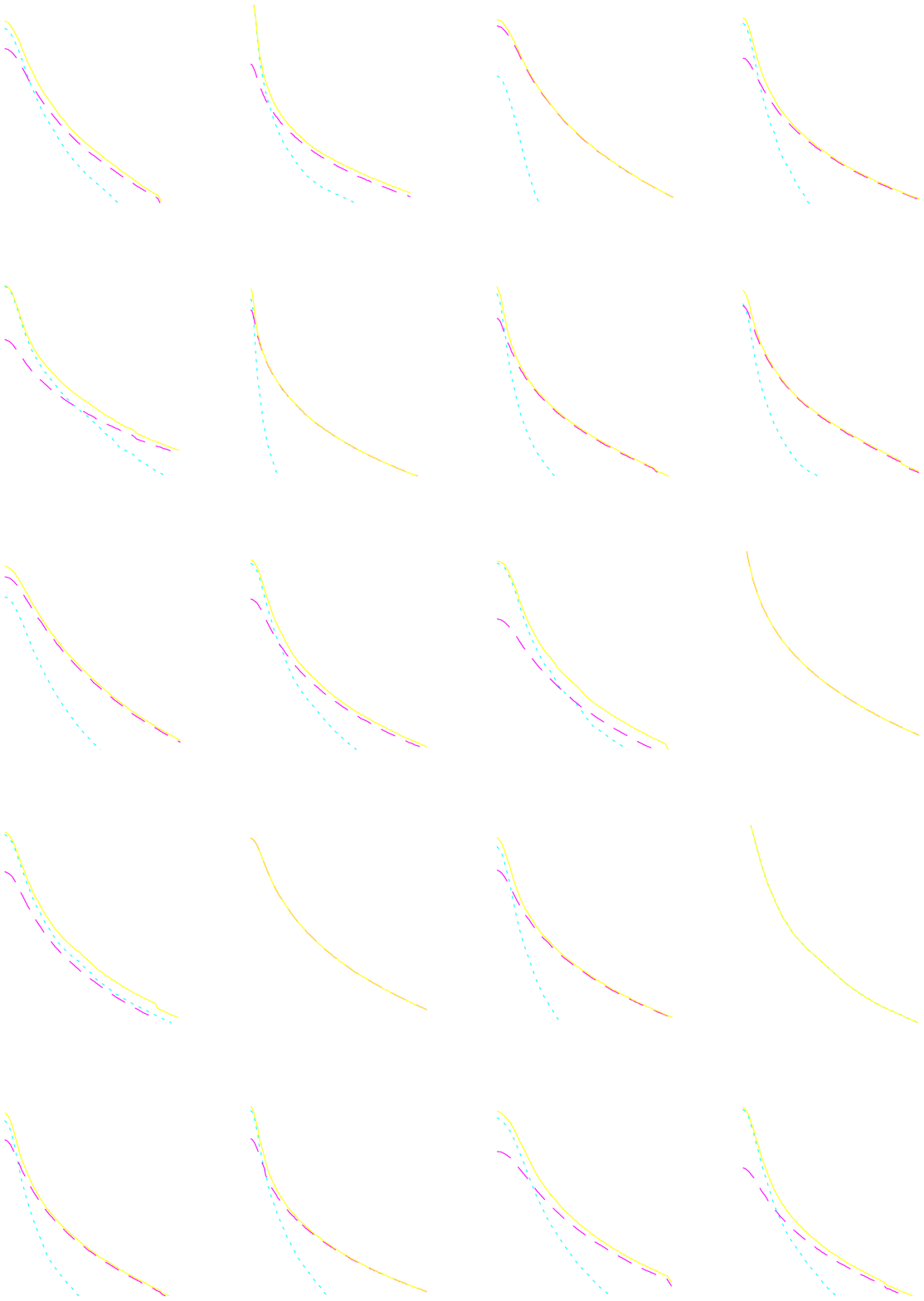,width=18cm,height=24cm}
\caption {The observed radial luminosity profiles of each BL Lac object 
(filled squares), superimposed to the fitted  model consisting of the PSF 
(short-dashed line), de Vaucouleurs bulge (medium-dashed line) and/or 
exponential disk (long-dashed line). The solid line shows the total model fit.
}
\end{figure*}

\begin{figure*}
\psfig{file=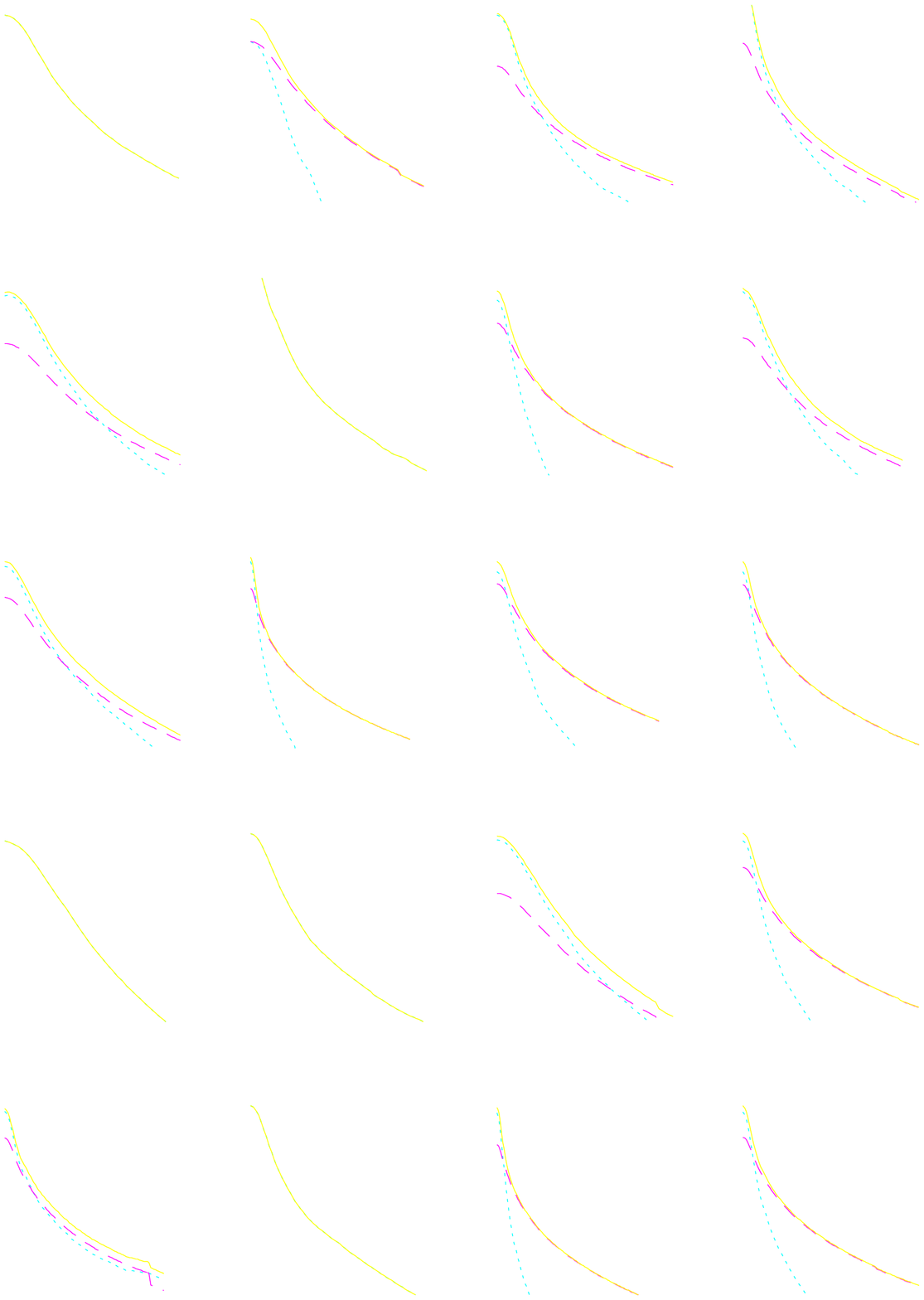,width=18cm,height=24cm}
{\bf Fig.~2.}~~Continued.
\end{figure*}

\begin{figure*}
\psfig{file=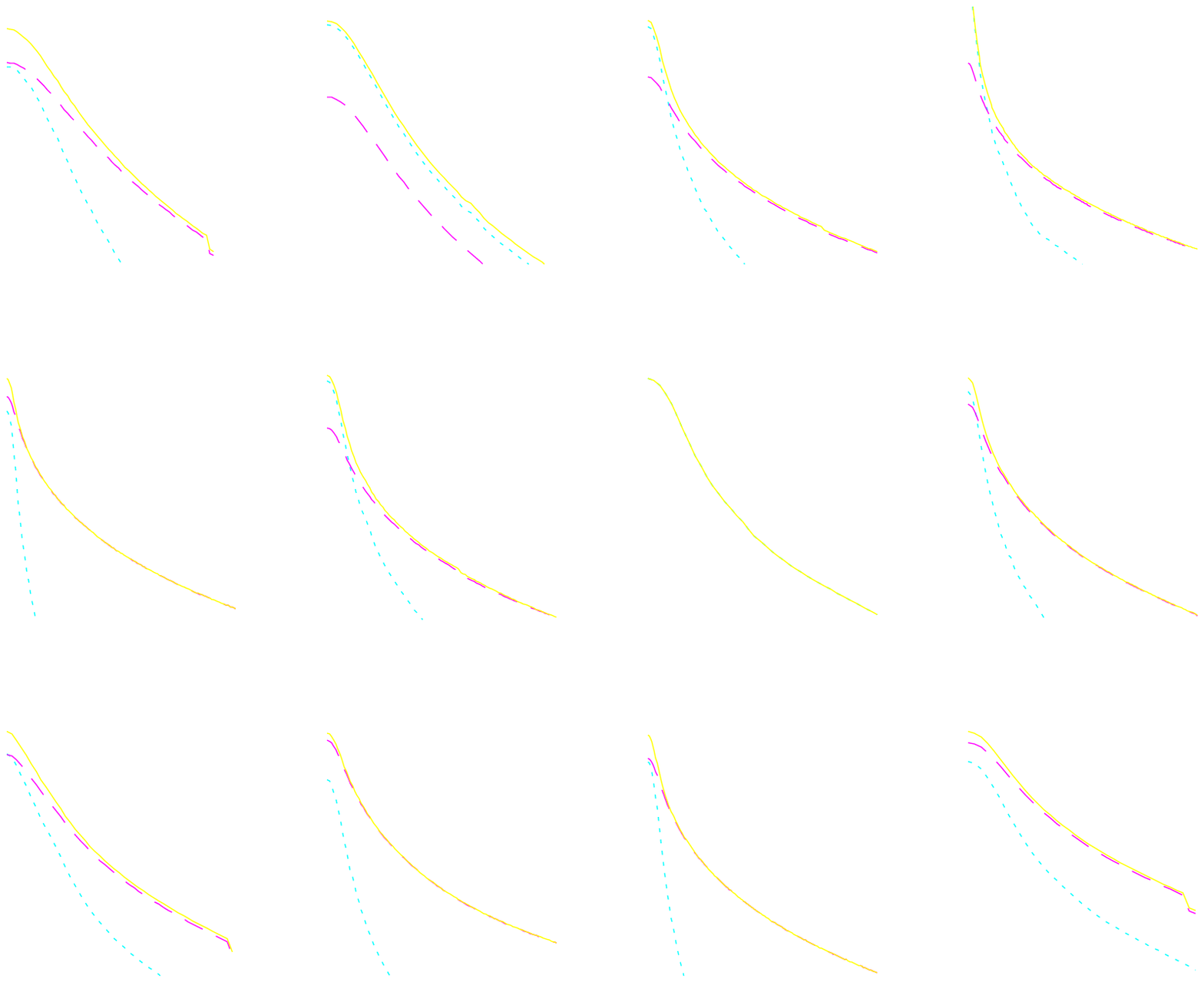,width=18cm,height=14cm}
{\bf Fig.~2.}~~Continued.
\end{figure*}

The absolute magnitudes of the nucleus and the host galaxies are reported in 
Table 3, columns (8)-(12), where 
column (8) gives the applied K--correction, columns (9) and (10) the 
absolute magnitudes of the nuclear component and the host galaxy, and columns 
(11) and (12) the scale length and  surface brightness at the effective 
radius.  The distributions of the host galaxy magnitude in the EMSS and Slew 
samples are illustrated in Fig. 3. The average absolute magnitude of the host 
galaxies is $<M_R(host)>$ = --23.85$\pm$0.59 (all), --23.94$\pm$0.50 (EMSS) 
and --23.74$\pm$0.68 (Slew), the average scale length of the host is 
$<$R(e)$>$ = 8.9$\pm$4.8 kpc (all), 8.8$\pm$5.2 kpc (EMSS) and 9.1$\pm$4.3 
kpc (Slew), while the average absolute magnitude of the nucleus is 
$<M_R(nucleus)>$ = --23.2$\pm$1.6 (all), --23.3$\pm$1.3 (EMSS) and 
--23.2$\pm$1.9 (Slew). The average host luminosity is in good agreement with 
previous studies (e.g. $<M_R>$ = --23.7$\pm$0.7, WSY96; 
$<M_R>$ = --23.7$\pm$0.6, U99b) and confirms that the host galaxies of BL 
Lacs are almost without exception giant ellipticals. No significant 
difference is found in the distribution of these values between the EMSS and 
Slew samples. At high luminosities the two distributions are very similar 
while at lower luminosities Slew hosts appear slightly but not significantly 
fainter than EMSS sources. This can be due differences in the selection 
procedure. A K-S test between the two distributions yields P$_{KS}$ = 0.1, 
confirming that the two distributions are practically indistinguishable.

\begin{figure}
\psfig{file=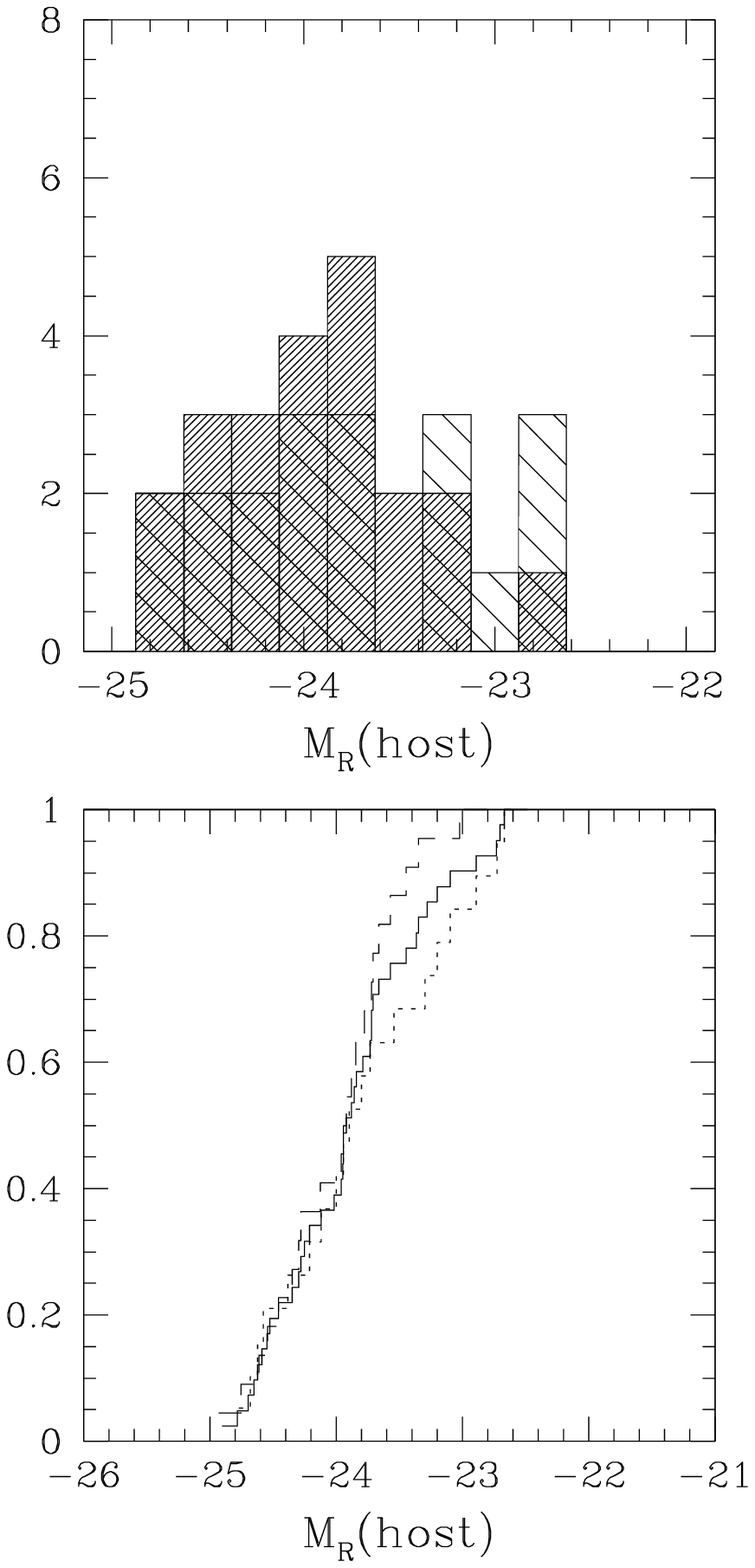,width=9cm,height=16cm}
\caption{{\it Top:} Histogram  of the absolute host galaxy 
magnitudes of the BL Lacs (thickly-hatched EMSS sample ;thinly-hatched
Slew sample). {\it Bottom:} Cumulative distributions for: All objects
(solid line), EMSS ( long-dashed line) and Slew (short-dashed line).
}
\end{figure}

\subsection{Comparison with previous observations of BL Lacs}

\subsubsection{Comparison with the CFHT survey}

Since we have 22 objects in common with the CFHT survey of BL Lac objects 
(WSY96). it is interesting to compare our results for individual sources with 
those obtained by WSY96. Of the 22 BL Lacs, three are unresolved by the CFHT 
and/or the NOT data. Thus, the final comparison is based on 19 resolved 
sources in common between the NOT and CFHT samples. Before comparing the 
results we have transformed the WSY96 host magnitudes from the Gunn $r$--band 
they used into our Cousins $R$--band assuming $r$--$R$ = 0.3 and applying a 
small correction ($\sim$ 0.1 mag) for the different cosmology used 
(q$_\circ$ = 0.5 instead of q$_\circ$ =0).  

Fig. 4 shows the apparent host magnitudes from the two studies plotted
against each other, with a one--to--one correspondence
superimposed. On average, the agreement is quite good, even if in few
cases the difference is quite large ( 1.7 mag for 0922+749 and 1.1 mag
for 0419).  For both these cases there are observations obtained with
HST (see the Appendix) that agree with our values within few tens of
mag.  The average and median difference in the host apparent magnitude is
$<m_{NOT} - m_{CFHT}>$ = --0.12$\pm$0.65 and 0.07, respectively.

\begin{figure}
\psfig{file=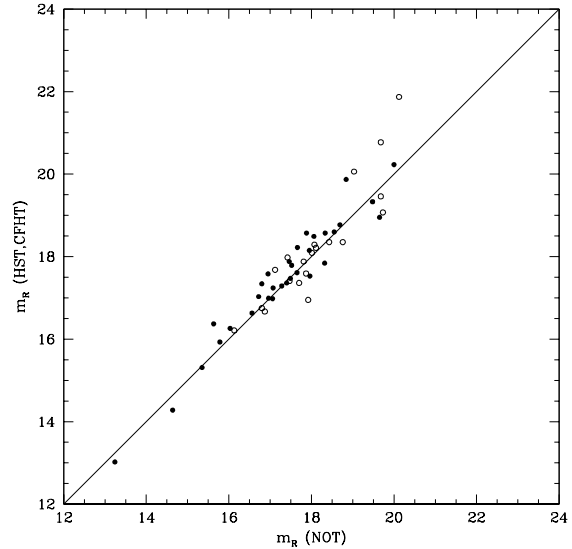,width=9cm,height=9cm}
\caption {Comparison of the apparent host galaxy magnitudes derived in 
this study for those BL Lacs in common with the CFHT survey (WSY96; 
open circles) and with the HST snapshot survey (U99b; full circles). 
A one--to-one correspondence is shown as a solid line.
}
\end{figure}

\subsubsection{Comparison with the HST snapshot survey}

We have 40 objects in common with the HST snapshot survey of BL Lac 
objects (Scarpa et al. 1999b; U99b). Although the NOT and HST data are taken 
with different instruments and have different spatial resolution, the 
data have been analyzed homogeneously, facilitating comparison between 
the samples. Of the 40 BL Lacs, four are unresolved by HST. Of these 
four sources, 1ES 0647+250, MS 1402.3+0416 and 1ES 1517+656 remain 
also unresolved by us, while for 1ES 0033+595 we are able to detect 
a probable nebulosity. This observation is however complicated  because 
of the presence of a very bright star close to the target (see individual notes). 

The  comparison is therefore based on 
35 sources in common between the NOT and HST samples. 

Fig. 4 shows the apparent host magnitudes from the two studies plotted
against each other, with a one--to--one correspondence superimposed.
The average and median difference in the host magnitude with respect
to the results of this study are $<m_{NOT} - m_{HST}>$ = 0.2$\pm$0.4
and 0.1, respectively.

In addition we show in Fig. 5 the comparison of the distributions of
the absolute host magnitudes from the two whole surveys. On average
the agreement is also very good.  The average host and nuclear
luminosities of the HST survey are $<M_R(host)>_{HST}$ =
--23.76$\pm$0.57 and $<M_R(nucleus)>_{HST}$ = --23.2$\pm$1.5. The
difference of average and median host luminosity with respect to the
results of this study are $<M_{NOT}> -- <M_{HST}>$ = -0.1 and -0.2,
respectively.

\begin{figure}
\psfig{file=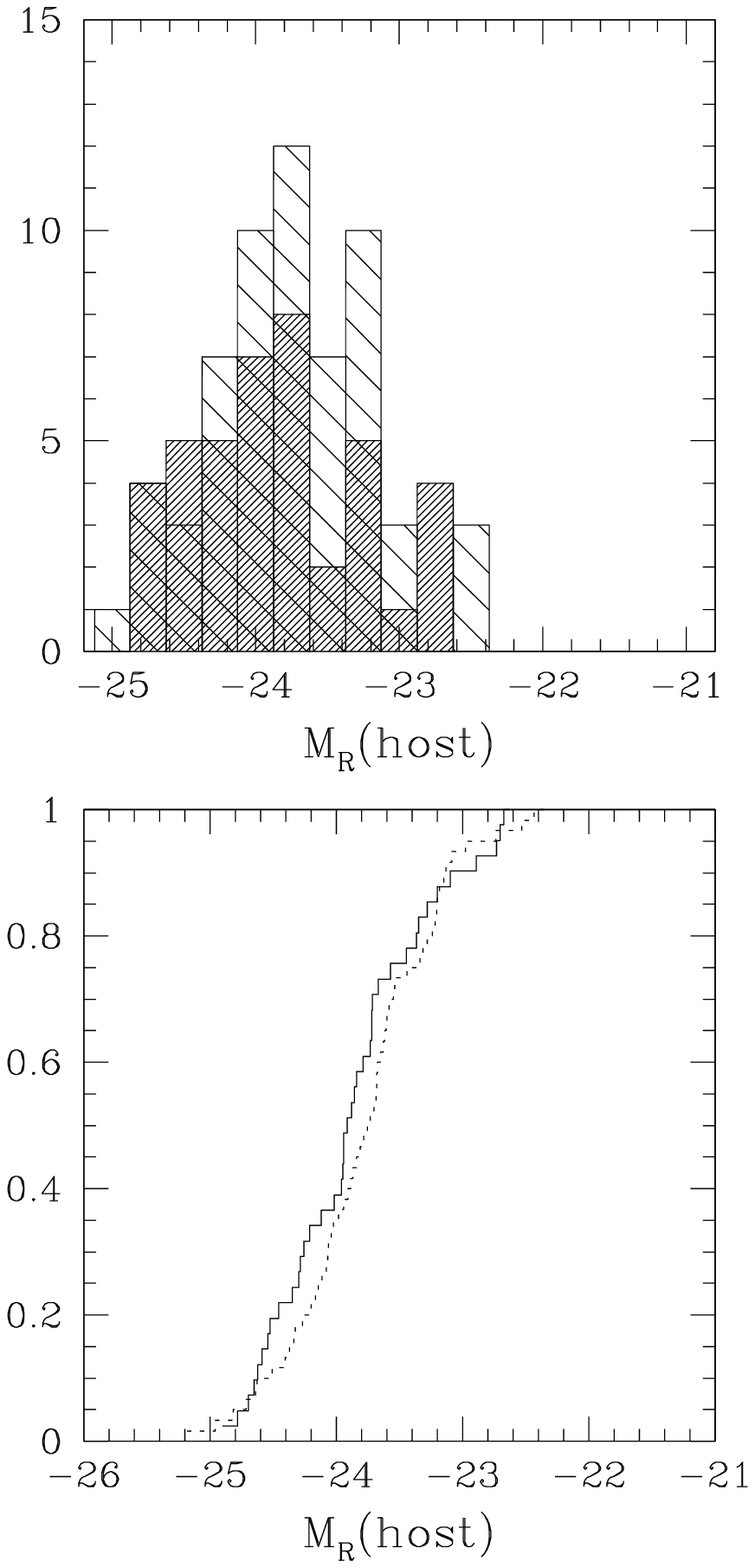,width=9cm,height=16cm}
\caption {Histogram and cumulative distribution of the absolute host 
galaxy magnitudes of the BL Lacs in this study (thickly hatched; solid 
line) and the whole HST snapshot survey (U99b; thinly hatched; short-dashed line.}
\end{figure}

In Fig. 6 we compare the radial profiles observed in this study of
three BL Lacs with those derived from HST images (Scarpa et al
1999b). The three objects were chosen to represent different well
resolved (MS 0257.9+3429), poorly resolved (1ES 1218+304) and
unresolved sources (MS 1402.3+0416) both in the NOT and HST
images. While the HST data allow one to investigate the host much
closer to the nucleus, the NOT images are clearly much deeper. This is
partly due to the longer exposure times and to the favorable pixel
scale of the NOT data. This translates into a better capability for
the NOT images of mapping the faint outer regions of the
galaxies. Also, Fig. 6 clearly shows the effect of different seeing on
the PSF in the ground based images (0\farcs5 for MS 0257.9+3429 and
1\farcs0 for 1ES 1218+304). An optimal characterization of the host
galaxies would thus be obtained by combining the high resolution (HST)
images with the deeper ground-based observations. The relevance of
this point in a larger sample of BL Lacs will be discussed in a
forthcoming paper.

\begin{figure}
\psfig{file=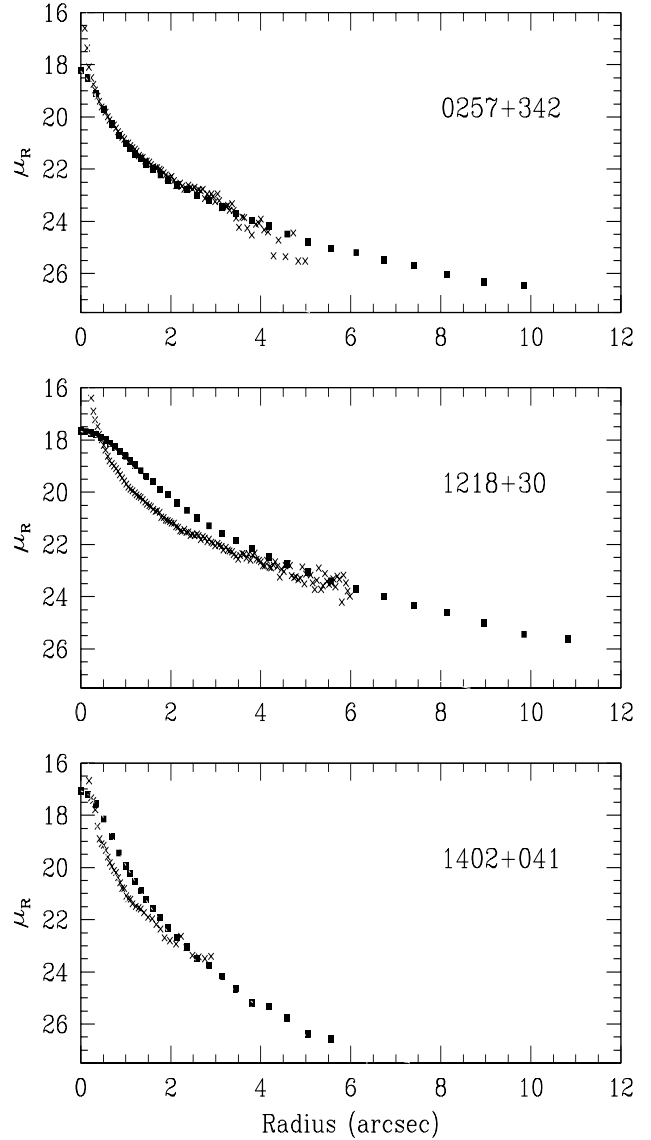,width=9cm,height=16cm}
\caption {Comparison of the radial luminosity profiles of selected BL 
Lacs derived from the NOT images (filled squares) and from the HST 
images (crosses). For comparison, a PSF profile matching the nucleus is 
plotted for both the NOT (dotted line) and the HST images (dashed 
line). Different representative cases are shown: a well resolved object 
(top panel), a poorly resolved object (middle panel) and an unresolved 
object (bottom panel).  }
\end{figure}

\subsection{Comparison with radio galaxies}

According to the unified schemes of radio-loud AGN (see e.g. Urry \& 
Padovani 1995), BL Lacs and RGs of F-R I type morphology are identical 
objects seen along different angles with respect to the relativistic jet. 
A number of optical studies have reported on the optical properties of RGs 
(see Govoni et al 1999 and references therein).  These have almost 
unanimously found that hosts are luminous ellipticals sometimes interacting 
with other galaxies, other times rather isolated. A general trend has 
been reported suggesting that F-R I RGs are on average brighter and 
in denser environments than those hosting F-R II RGs. Comparison among 
various samples is hindered by the selection procedure and by possible 
systematic effects introduced by the analysis (e.g. extinction,  
K-correction, passband, method of measurement of galaxy magnitude, etc). 
In order to reduce as much as possible the systematic effects we compare 
our results with those obtained by Govoni et al (1999) for a large sample 
of low redshift RGs. 
This sample includes 79 RGs of both F-R I and F-R II 
type in the redshift interval 0.01 to 0.1. $R$--band imaging is used to 
investigate in detail the morphological and photometric 
properties of the radio galaxies. 
This includes the analysis of the luminosity profile using the 
same procedure and the corrections used in this work.

The average host luminosity of the F-R I and F-R II RG hosts are
$<M_{FRI}>$ = -24.1$\pm$0.6 and $<M_{FRII}>$ = -23.6$\pm$0.7. The BL
Lac hosts appear therefore on average slightly brighter than F-R II
hosts, but also slightly fainter than the F-R I RG hosts, which are in
turn quite similar to BCGs in moderately rich clusters at z$<$0.15
($<M_{BCG}>$ = --24.1$\pm$0.3; Hoessel et al. 1980).  Note that since
the RGs are at lower redshift, any cosmological evolutionary
correction makes the difference from FR I even larger.  A similar
trend was also noted by Lamer Mc Hardy Newsam (1999) comparing BL Lacs
and FR I galaxies in the near infrared.

On the other hand the comparison between BL Lac hosts and the whole
sample of RG shows a general good agreement.  In Fig. 7 we report the
histogram and cumulative distribution of the absolute host magnitudes
of the BL Lacs and the RGs from Govoni et al. (1999).  The two
distributions are rather similar with only a significant excess of
more bright RG that are not present in the BL Lacs.  If we compare BL
Lacs with F-R I and F-R II separately the agreement is formally good
for F-R II but not for F-R I.  A K-S test yields formally P$_{KS}$ =
0.250 and 0.004 for F-R II and F-R I respectively compared with BL Lac
host luminosities.

\begin{figure}
\psfig{file=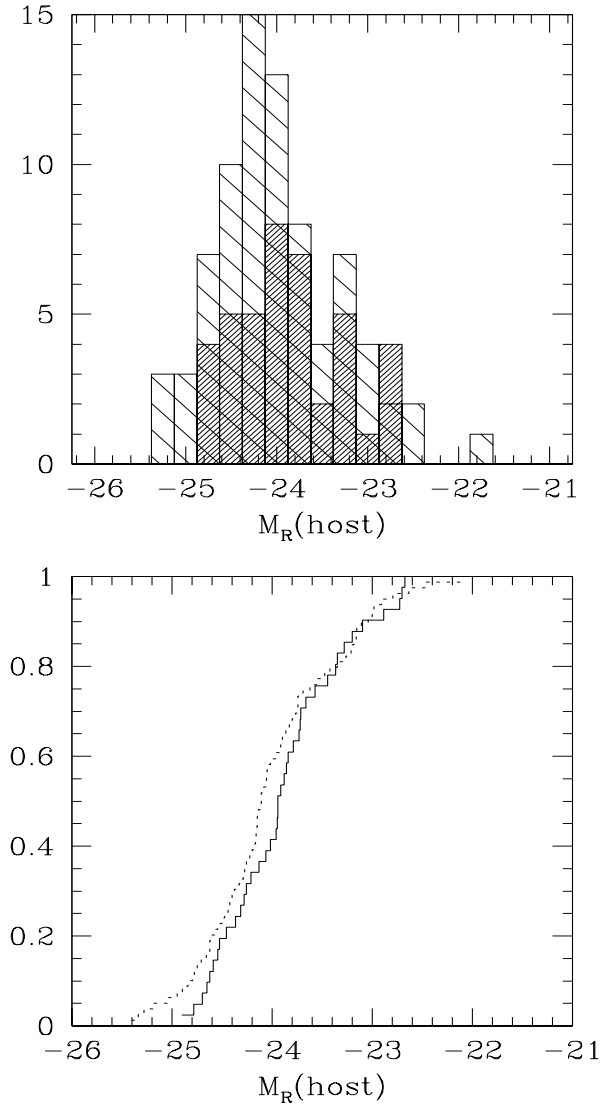,width=9cm,height=16cm}
\caption {Histogram and cumulative distribution of the absolute host 
galaxy magnitudes of the BL Lacs in this study (thickly hatched; solid 
line) and the RGs (Govoni et al. 1999; thinly hatched; short-dashed line).  }
\end{figure}
Since the selection procedure that identifies samples of  BL Lacs and of RG 
is different, we cannot rule out that selection effects may introduce some 
bias. These effects should be related to correlations between host 
optical luminosity and X-ray and radio properties which are used to 
classify the objects. 

The present result indicates that from the point of view of host 
luminosities either the parents of BL Lacs are RG of both types (I and II) 
or for some reason brightest F-R I must be excluded from the parent 
population as proposed by WSY96. 

The first association of BL Lacs with F-R I was done on the basis of 
similar extended radio emission (Browne 1983, Wardle et al 1984; Antonucci 
\& Ulvestad 1985). There are however  some observations 
(e.g. Kollgaard et al 1992, Murphy et al 1993) that indicate that 
at least some BL Lac object may have diffuse emission more similar to that 
of F-R II than F-R I.  The same conclusion is reached from the kpc 
scale radio polarization study of six BL Lacs (Stanghellini et al 1997) 
where the magnetic field is parallel to the radio jet axis for most of
its length as found in 
F-R II sources. If the radio morphology of these BL Lacs be of F-R I type 
the  magnetic field should be dominated by the component perpendicular to jet axis.

Our results therefore are consistent with the idea that BL Lacs avoid
 the BCGs in rich clusters at low redshift (Wurtz et al.1997) although
 there are examples of BL Lacs in very luminous hosts and members of
 group/clusters of galaxies (see 1ES 0414+009 and PKS 0548-322; this
 study and  Falomo et al 1995; 1Es1741+196 Heidt et al 1999).  This idea is
 also supported by the narrow-angle tail radio morphology seen in many
 BL Lacs and non-BCGs (e.g. Owen \& Laing 1989; Perlman \& Stocke
 1993) and by the correlation between cluster richness and the BCG and
 BL Lac luminosity (Thuan \& Romanishin 1981; Wurtz et al. 1997).

On the other hand evidence is growing that also from the point of view
of the extended radio luminosities many BL Lacs are quite different
from FR I and share the characteristic luminosity of FR II sources
(Cassaro et al 1999). This evidence together with our results on the
host luminosities led to argue that both types of RG form the parent
population of BL Lacs.

\subsection{The Fundamental Plane and Hubble diagram}

It is well established that elliptical galaxies form families of 
homologous systems with characteristic parameters  R(e), $\mu$(e) 
and velocity dispersion $\sigma$. These are commonly represented in 
the Fundamental Plane (see e.g. Djorgovski \& Davis 1987). We 
have investigated the properties of the BL Lac hosts in the 
projected Fundamental Plane (F-P) concerning the central surface 
brightness $\mu$(e) and the scale length R(e). Surface brightness 
$\mu$(e) derived from the fit (see Table 3) was corrected for 
Galactic extinction, K--correction and for the (1+z)$^4$ 
cosmological dimming. Fig. 8 shows the correlation between $\mu$(e) and 
log R(e) for the BL Lac hosts and the low redshift RG hosts (Govoni et 
al. 1999). It can be seen that the behavior of the EMSS and Slew BL Lac 
hosts are similar. Both BL Lac hosts and the RG hosts (Govoni et al. 
1999) follow the Kormendy relation for giant massive ellipticals 
(e.g. Capaccioli et al. 1992), with a best-fit correlation $\mu$(e) = 16.45 
+ 4.6 $\times$ (log R(e)). Practically no host galaxy is in the 
(scatter) area at log R(e) $<$ 0.5 kpc. This confirms that the BL Lac 
hosts are almost exclusively drawn from the population of giant 
ellipticals and not from normal field ellipticals. Note that the BL Lac 
hosts seem to avoid the area of the brightest and largest galaxies in 
the bottom right hand corner of Fig. 8, similarly to the result based on 
the total host luminosities (Fig. 7). 

\begin{figure}
\psfig{file=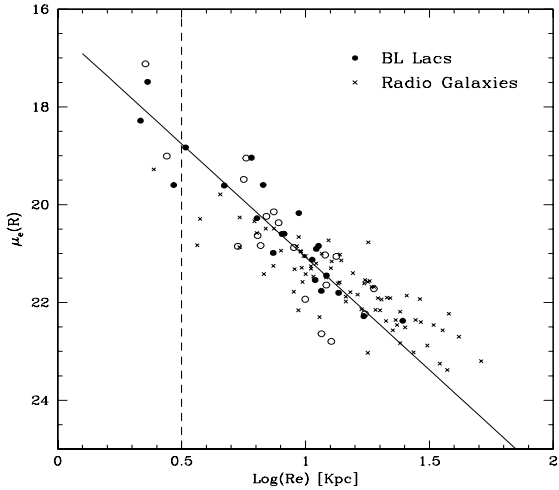,width=9cm,height=8cm}
\caption {Kormendy relation for BL Lac host galaxies: 
EMSS (full 
circles)and Slew (open circles) objects are compared with RG hosts (Govoni et 
al. 1999; crosses). The solid line shows the best fit, while the dashed line 
shows the dividing line between normal and giant ellipticals (Capaccioli 
et al. 1992).
 }
\end{figure}

In Fig. 9 we show the apparent R magnitude vs. redshift Hubble diagram 
for the BL Lac hosts (this work), together with data for RGs (Govoni et 
al. 1999), compared with 
the expected relationship  for elliptical galaxies derived from 
passive stellar evolution models (Bressan et al 1994) 
normalized to the average redshift and magnitude of the low redshift RGs 
from Govoni et al. 1999). The resolved BL Lac hosts lie within about 1 
mag spread on the H--z relation, as do the RGs. Based on this diagram we 
can estimate the redshift of the four objects with unknown redshift 
but resolved in our images. These are 0033+59, 1106+24, 2037+52, and 
2336+05 (see Fig. 9)  for which we derive a Hubble law redshift of 
0.43, 0.58, 0.14 and 0.40 respectively.

\begin{figure}
\psfig{file=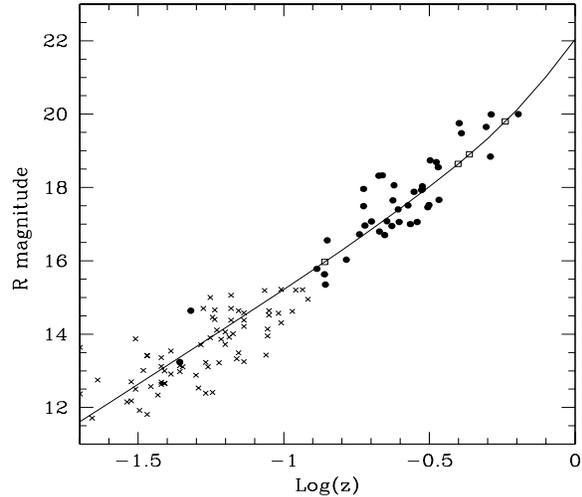,width=9cm,height=8cm}
\caption {Hubble diagram for BL Lac hosts (filled circles) and RGs 
(crosses) by Govoni et al. 1999. 
The solid line shows the expected behavior for a massive elliptical galaxy 
undergoing  passive stellar evolution (Bressan et al 1994). 
The expected position on the Hubble diagram for the four resolved BL Lacs 
with unknown redshift are marked with open squares (see text).
 }
\end{figure}

\subsection{Luminosities of the host and the nucleus}

Optical images of BL Lacs are characterized by the superposition of 
two components: the extended starlight emission from the galaxy and 
the non-thermal unresolved  bright nucleus. We discuss here the properties 
of these two components as derived from our sample.

Fig. 10 shows the relationship between absolute magnitude of hosts 
and redshift of our BL Lacs and of  other samples of AGN. All BL Lac hosts 
at all redshifts have luminosities between that of a passively evolving 
M$^*$ and M$^*$--2 mag galaxies, where M$^*$ is the characteristic 
luminosity of a Shechter luminosity function. While there are RG hosts (Govoni 
et al. 1999) with  luminosities larger than M$^*$--2, none of the observed 
BL Lacs are found in such a luminous galaxy. Note that any correction due 
to different average redshift would make this difference even larger. 
There is a suggestion of a positive correlation of host luminosity 
with redshift, consistently with what is expected from passive 
stellar evolution models for elliptical galaxies (e.g. Bressan et al. 
1994; Fukugita et al. 1995), and the evolution of galaxies in 
clusters (Ellingson et al  1991), although the scatter is large. 
This trend is similar to that suggested by WSY96 and is consistent with 
the average value found for higher redshift flat spectrum radio quasar 
(FSRQ) host galaxies (Kotilainen et al 1998a): $<z>_{FSRQ}$ 
= 0.673, $<M_{host}>_{FSRQ}$ = --25.3 and even with high redshift 
RLQs (Lehnert et al 1992): $<z>_{RLQ}$ = 2.34, $<M_{host}>_{RLQ}$ = --26.3. 

\begin{figure}
\psfig{file=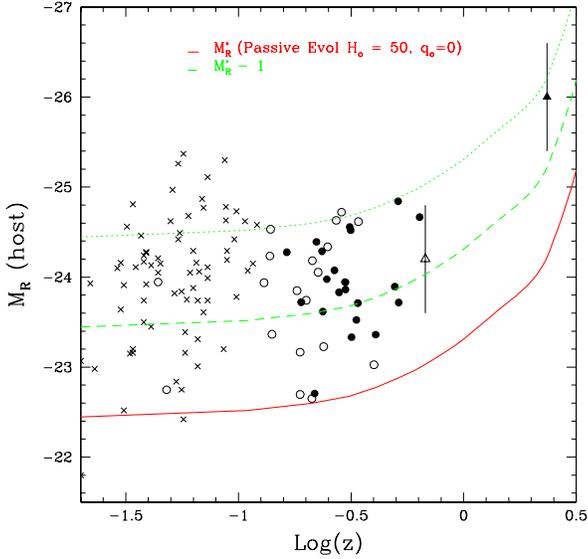,width=9cm,height=9cm}
\caption { Correlation between the absolute host magnitude of the 
BL Lacs in EMSS (full circles) and Slew (open circles) with redshift. Solid, 
dashed and dotted lines show a passively evolving M$^*$, M$^*$--1 and 
M$^*$--2 mag galaxy, respectively (Bressan et al 1994 ). Also plotted are RGs 
from Govoni et al 1999 (crosses) and the average values for FSRQs (open 
triangles; Kotilainen et al. 1998a) and high z RLQs (filled triangles; 
Lehnert et al. 1992).
}
\end{figure}

Fig. 11 shows the histogram of the ratio between nuclear and host 
luminosity for the resolved BL Lac objects. This ratio ranges from 0.03 to 
3, with average $\sim$0.8 and median $\sim$0.4. Interestingly, we find 
an apparently bimodal distribution with two peaks around 
log(LN/LH) $\sim$--0.6 and log(LN/LH) $\sim$0.2. This behavior is present 
in both the EMSS and Slew sub-samples. 

While the detected large range in the luminosity ratio within
individual objects can be due to differences in the intrinsic nuclear
or host luminosity, or a difference in the beaming factor from one
object to another, it is difficult to explain the apparent bimodality
present in both EMSS and Slew samples.

In comparison, the luminosity ratio in the $V$--band for the low 
redshift quasars studied by Bahcall et al. (1997) is $\sim$10 while for 
the RGs observed in the $R$--band (Govoni et al 1999) this ratio is 
less than 0.1. For BL Lacs therefore the observed nuclear optical luminosity 
seems intermediate between that of RG and quasars.

\begin{figure}
\psfig{file=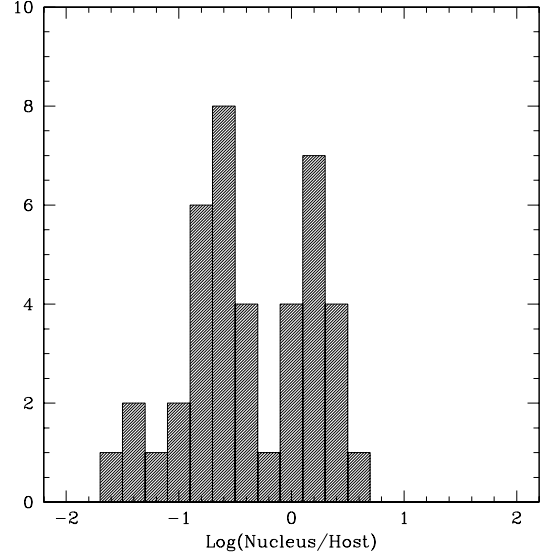,width=9cm,height=9cm}
\caption { Histograms of the ratio between nuclear and host 
luminosity for resolved BL Lac objects. 
}
\end{figure}


In order to investigate the relationship of the optical nuclear 
luminosity with that in different bands we have performed a 
partial correlation analysis among radio, optical, and X-ray luminosities. 
We used data in Table 1 together with nuclear optical data of Table 3 
and used the procedure described in Padovani (1992).  For all objects 
with known redshift we computed the Spearman rank order 
correlation coefficients between luminosities and between luminosity in 
a given band and redshift.  The results of the correlation analysis are 
given in Table 4.  It turns out that apart the weak correlation with 
redshift (which is marginally significant only for L$_x$) the 
only significant correlation among luminosities after subtraction of 
the redshift dependency is between L$_o$ and L$_r$.  A similar result 
was obtained from the analysis of a smaller sample of X-ray BL Lacs 
by Padovani (1992) but with a smaller correlation coefficient.

\begin{table}
\begin{center}
\caption{Correlation coefficients for BL Lacs luminosities.}
\begin{tabular}{lllll}
\multicolumn{5}{c}{}\\
\hline
     & L$_r$ &L$_o$& L$_x$ & z \\
L$_r$ & 1.0 & 0.74 & 0.55 & 0.30 \\
L$_o$ & ... & 1.0 & 0.48 & 0.27 \\
L$_x$ & ... & ... & 1.0 & 0.45 \\
 z    & ... & ... & ... & 1.0 \\
\hline
\end{tabular}
\end{center}
\end{table}

Fig. 12 shows the relationship between the absolute host and 
nuclear magnitudes of the BL Lacs in EMSS and Slew. There is an indication 
of a weak correlation (Spear correlation coefficient. = 0.3) between the 
two parameters, in the sense that more luminous nuclei reside in 
more luminous hosts. 

However, we note the obvious selection effects that may de-populate the
upper left hand (faint nuclei in luminous hosts) and lower right hand
(faint hosts swamped by luminous nuclei) corners of the diagram.
While the first effect could bias the original samples (see
e.g. Browne \& Marcha 1993) the second should have marginal effect
since, excluding misclassified objects we are able to resolve 90\% of
observed targets.

The putative correlation is, however, consistent with the 
luminosity/host-mass limit  found when considering AGN samples at higher 
redshift and with more luminous nuclei and hosts (Kotilainen et al. 
1998a; Lehnert et al. 1992). Also, assuming that BL Lac activity results 
from accretion of material onto a super-massive black hole, it is in 
agreement with the relationship found by Magorrian et al. (1998) from 
HST kinematic study between the mass (luminosity) of the black hole and 
the mass (luminosity) of the spheroid component in nearby galaxies. McLure 
et al. (1999) detected a similar weak correlation for a small  sample of 9 
RLQs at z$<$0.35, and calculated that most of their RLQs radiate at a 
few percent of the Eddington luminosity. The correlations found in this 
study and in McLure et al. (1999) indicate that the Magorrian et 
al. relationship extends to galaxy and host galaxy masses at 
cosmological distances. On the other hand the lack of a strong 
correlation found in this 
study of relatively nearby and modest luminosity AGN may indicate that 
the onset of the correlation occurs only after a certain level in 
nuclear and/or galaxy luminosity has been reached (c.f. 
with Kotilainen et al. 1998a). 

\begin{figure}
\psfig{file=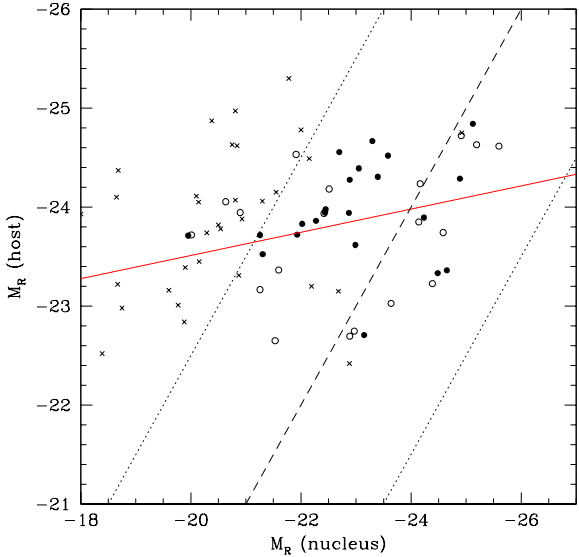,width=9cm,height=9cm}
\caption {  Absolute host galaxy magnitude plotted against
 absolute nuclear magnitude of the resolved BL Lacs: EMSS (full
circles) and Slew (open circles).  Also plotted are values for low redshift 
RGs (crosses) by Govoni et al. Solid line is a simple regression
fit to the all data.  Dashed line is the one to one 
relationship while dotted lines represent
loci in the diagram with luminosity ratio = 0.1 and 10.
}
\end{figure}

\section{Conclusions}

We have presented high resolution homogeneous optical observations of
a large data set of BL Lac objects drawn from two complete samples of
X--ray selected (high frequency peaked) sources.  We are able to
resolve $\sim$ 90\% of the observed targets and study the properties
of their host galaxies.  It turns out that all galaxies are well
represented by elliptical model with mean luminosity $<M_R(host)>$ =
--23.9 and an observed average nuclear source-to-host luminosity ratio
of $\sim$ 1. It is also shown that hosts of BL Lacs are almost
exclusively drawn from the population of massive elliptical galaxies.
No cases of disc dominated systems hosting BL Lacs are found
supporting the view that all radio-loud active galaxies are dominated
by the spheroidal component.

The global properties of the galaxies hosting BL Lacs follow the same
relationships of normal (non active) giant ellipticals under passive
stellar evolution.  The comparison of properties of BL Lac hosts with
those of low redshift radio galaxies indicates that from the point of
view of galaxy luminosity both FR I and FR II radio galaxies form the
parent population of BL Lacs. A result that is supported recent radio
imaging and polarization studies that show many BL Lacs exhibiting the
characteristics of FR II sources.

\appendix

\section{Notes on individual objects}

\subsection{Unresolved BL Lacs}

In spite of the good seeing conditions seven BL Lacs remained unresolved. 
For all these objects the redshift is unknown or uncertain. For two 
unresolved BL Lacs (MS 0950.9+4929 and 1ES 1517+656), the redshift (or 
lower limit of redshift) is high enough to explain the lack of detection of 
the host galaxy. For the other five  unresolved objects (1ES 0647+250, 
1ES 1118+424, MS 1256.3+0151 MS 1402.3+0416 and 1ES 2321+419), there are  
only tentative  redshifts  which are, however, inconsistent with the 
results from our images.

We determined a lower limit of the redshift assuming these objects 
are surrounded by a typical  elliptical host galaxy as found in this study 
(M(R) = --23.8 and R(e) = 10 kpc). A simulated galaxy was produced 
and superposed onto  the image of the observed sources assuming 
various redshifts.  This produced images of the object that appeared 
resolved or not depending on the redshift used. A lower limit of the 
redshift was set when  the galaxy become  undetectable. Of course using 
host galaxies that are less luminous and smaller would make these limits lower.

The derived lower limits for the unresolved BL Lacs are: z $>$ 0.3 for 
1ES 0647+250, z $>$ 0.6 for MS 0950.9+4929, z $>$ 0.5 for 1ES 1118+424, 
z $>$ 0.65 for MS 1256.3+0151, z $>$ 0.5 for MS 1402.3+0416, z $>$ 0.45 
for 1ES 1517+656 and z $>$ 0.45 for 1ES 2321+419. With the exception of 
1ES 0647+250, the observed  magnitude of these objects is consistent with 
them being at moderately high redshift, thus remaining unresolved. The case 
of 1ES 0647+250 could be due to either a very bright (or beamed) nucleus or 
an under-luminous host galaxy. Since some of these BL Lacs  have  
tentative redshift (see individual notes) either these objects  are hosted 
by an atypically faint host galaxy or, more likely, the  redshift is wrong. 

\noindent{\bf 1ES 0647+250}: A tentative redshift z = 0.203 has been 
derived for this BL Lac, which is unresolved in the NOT images taken 
with very good seeing (FWHM 0\farcs65). It is also  unresolved in the 
HST snapshot survey (Scarpa et al. 1999b), who derived an upper limit 
m(host) $>$19.1. At z = 0.203, this corresponds to M(host) $>$--21.7, i.e. 2 
magnitudes fainter than an average BL Lac host. 

\noindent{\bf MS 0950.9+4929}: This BL Lac, for which a lower limit to 
the redshift of z $>$ 0.5 has been derived (Perlman, priv. comm.), 
remains unresolved in the NOT data, even in good seeing. It was 
also unresolved by WSY96, who derived an upper limit m(host) $>$19.2. 
The corresponding absolute luminosity limit from WSY96 is M(host) $>$ 
--24.4, in good agreement with the average value of the BL Lac hosts 
derived here. 

\noindent{\bf 1ES 1118+424}: This BL Lac, at an uncertain redshift z = 
0.124 (Perlman et al, priv. comm), remains unresolved in the NOT data, 
even in very good seeing (FWHM 0\farcs60  ). The redshift is inconsistent 
with results from our image. 

\noindent{\bf MS 1256.3+0151}: This BL Lac, at tentative redshift z = 
0.162, is unresolved in the NOT data but the seeing during this 
observation was rather poor (FWHM 1\farcs34 ).  However, even taking 
into account the poor seeing, if it is at z = 0.162 its host galaxy should 
be easily detectable (unless it is atypically faint). We estimate it 
should be at $z >$ 0.5 if hosted by a standard galaxy.

\noindent{\bf MS 1402.3+0416}: This BL Lac, at a tentative redshift z 
= 0.344, remains unresolved in the NOT data.
It was also unresolved by WSY96, who derived an upper limit m(host) 
$>$18.4, which at z = 0.344 corresponds to M(host) $>$ --23.9. Similarly, 
it remained unresolved in the HST snapshot survey (Scarpa et al 1999b), 
who derived an upper limit m(host) $>$19.4, which at z = 0.344 corresponds 
to M(host) $>$ --22.9. These upper limits still allow for the existence of 
a typical elliptical host galaxy which at the moderately high 
redshift remains unresolved. 

\noindent{\bf 1ES 1517+656}: This BL Lac, for which Beckmann (priv. 
comm.) has derived a spectroscopic lower limit  of z $>$0.7 based on 
the presence of MgII and FeII absorption lines, remains unresolved in the 
NOT data, even in good seeing (FWHM 0\farcs70  ). It was also unresolved in 
the HST snapshot survey (Scarpa et al 1999b), with an upper limit 
m(host) $>$19.9, which at z $>$0.7 corresponds to M(host) $<$ --25.2.  
At this upper limit the observations are consistent with the presence of 
a luminous host galaxy.  At HST resolution, this BL Lac shows an 
unusual morphology with three non homogeneous arclets surrounding the 
point source at 1-2 arcsec of distance (Scarpa et al 1999a). We are able 
to see clearly the more external arc and can detect the two more 
internal features after subtraction of a scaled PSF. Our deeper images 
show that at least the most external ring does not extend more than what 
it is seen on HST data. Moreover we detect many other faint sources in 
the immediate 5 arcsec environment.

It was proposed that these arcs are effects of gravitational lensing 
produced by a foreground galaxy. Our image is much deeper than that 
obtained with HST but no  signature of foreground galaxy is found. 
This therefore weakens the lens hypothesis.

\noindent{\bf 1ES 2321+419}: This BL Lac, at a tentative redshift z = 
0.059, remains unresolved by our images. Since the seeing conditions 
were very good (FWHM 0\farcs69  ) we believe the correct redshift 
is considerably larger than that proposed.

\subsection{Misclassified BL Lacs?}

\noindent{\bf 1ES 0446+449}: This BL Lac at a tentative redshift z = 0.203 
was resolved by the HST snapshot survey (Scarpa et al. 1999b), but no 
point source was detected while a pure exponential (disk) brightness 
profile was observed.  The NOT data confirm that no point source is present 
in this object. Instead, we detect a very luminous  galaxy with m(host) = 
15.1, corresponding to an extremely luminous galaxy (M$_R$ = --29.1 at 
the proposed redshift) making it a very unusual source. 
However, a more likely explanation is that the correct redshift for this 
object is much lower than z = 0.203 and that this source is not a BL Lac 
object. From re-inspection of the optical spectrum for this object (Perlman 
et al. 1996), we note that the redshift determination is probably wrong and 
the correct redshift is z$\sim$0.02, which would make the identified 
counterpart to be a normal low redshift elliptical galaxy, and not a BL 
Lac object. 

\noindent{\bf 1ES 0525+713}: The NOT data failed to detect 
a point source at the tentative redshift z = 0.249, while the elliptical host galaxy has m(host) = 17.7 and 
M(host) = --24.3.
Similar results was obtained from HST images (Scarpa et al 
1999b). 
Imaging results together with  the lack  a power law continuum in 
the optical spectrum (Perlman et al. 1996) strongly suggest it is not a BL 
Lac source.

\subsection{Resolved BL Lacs}

\noindent{\bf 1ES 0033+595}: The HST observations of this BL Lac at 
unknown redshift (although a tentative redshift z = 0.086 was derived 
by Perlman, priv. comm.) have been discussed by Scarpa et al (1999a). It is 
a gravitational lens candidate consisting of two objects (A and B; see 
Scarpa et al 1999a) with similar $R$--band brightness and separation of 
$\sim$1\farcs6. In addition to $R$--band images we obtained also data in 
the $U$--, $B$-- and $V$--bands. This allows us to compare colors of the 
two objects. These turned out to be significantly different with $U$--$B$ 
= 0.4 and -0.1, $B$--$V$ = 1.4 and 1.7, and $V$--$R$ = 0.9 and 1.9 for A 
and B, respectively. Due to the bluer color it is likely that the BL Lac 
is the object B while A is a red galactic star (the source is close to 
the galactic plane). 

The strong color differences strongly argue against the lens hypothesis. 
The most likely explanation is a chance alignment with a foreground star.

The probable BL Lac in the pair (component B) remained unresolved in
the HST snapshot survey (Scarpa et al 1999b), with an upper limit
m(host) $>$20.0. From the analysis of the NOT image we are able to
detect an excess of light at radii larger than 2 arcsec corresponding
to surface brightness fainter than 24 mag/arcsec$^{-2}$.  If the
nebulosity is attributed to the host galaxy this would correspond to
an object of 19.0 mag.  A caution is however needed for this detection
because of the presence of a very bright star in the field that
contaminates the background emission.

\noindent{\bf MS 0205.7+3509}:  This source, for which WSY96 preferred a 
disk host galaxy surrounding a significantly de-centered nucleus, was 
recently studied in detail by us (Falomo et al. 1997). We refuted both the 
disk model and the gravitational lens interpretation (Stocke et al. 1995), 
and identified MS 0205.7+3509 as an elliptical host galaxy with no 
de-centering but with a close companion galaxy.

\noindent{\bf MS 0317.0+1834}: WSY96 could not distinguish between 
elliptical and disk host for this source, for which they derived M(R) = 
--23.0 and R(e) = 5.1 kpc. After masking out the close companion, we 
clearly prefer an elliptical host with M(R) = --23.7 and R(e) = 4.7 kpc, 
a somewhat brighter host. 

\noindent{\bf 1ES 0347-121}: The host galaxy is clearly resolved with 
the radial profile well described by an elliptical model. We derive for 
the host M(host) = --23.2, identical to that derived by U99b. There is 
an interacting system of three galaxies located 12 arcsec N of the BL Lac 
but no signs of a physical connection with the BL Lac are apparent.

\noindent{\bf 1ES 0414+009}: This BL Lac was resolved into an elliptical 
galaxy with M(R) = --24.8 by the HST (U99b). It was also studied by 
Falomo (1996) who derived for the host M(R) = --24.3 and R(e) = 5 kpc. 
We derive for the elliptical host at z = 0.287 M(R) = --24.7 and R(e) = 5.8 
kpc, in good agreement with the other measurements. This source was 
also studied by WSY96, who could not distinguish between elliptical and 
disk morphology. They derived for the host M(R) = --24.4 and R(e) = 30.8 
kpc. While the host luminosity is in agreement, the scale length from WSY96 
is clearly too large. Note that 1ES 0414+009 is the dominant member of 
a moderately rich cluster at z = 0.287 (Abell class 0; McHardy \ea 1992; 
Falomo et al 1993).

\noindent{\bf MS 0419.3+1943}: WSY96 derived for the probably elliptical 
host M(R) = --23.5 and R(e) = 22.9 kpc, while U99b derived for the 
elliptical host M(R) = --24.0, and we identify the host as clearly an 
elliptical with M(R) = --24.8 and R(e) = 6.0 kpc. 

\noindent{\bf 1ES 0502+675}: The HST observations of this BL Lac at z = 
0.341 have been discussed by Scarpa et al (1999a). It is a
gravitational lens candidate with a double source of similar magnitude
with a separation of only $\sim$0\farcs3. Scarpa et al (1999a) found
the host galaxy has M(host) = --23.9. Using the NOT data we derive for
the host galaxy M(host) = --24.6.  The difference is likely due to the
fact that our image is much deeper than the one obtained with HST
reaching $\mu_R$ = 26.5 at 6 arcsec from the center compared with
$\mu_R$ = 24 at 2 arcsec from HST image.

\noindent{\bf MS 0607.9+7108}: For this BL Lac at z = 0.267, the morphology 
of its host galaxy has been controversial, mainly due to the presence
of a bright nearby star. WSY96 preferred a disk model (although they
could not rule out elliptical host) and derived M(R) = --24.8 and R(e)
= 8.9 kpc, while HST (U99b) indicated an elliptical host with M(host)
= --24.3.  Because of the presence of the close bright star we have
first subtracted a scaled PSF from the image removing the contribution
from the star and then extracted the brightness profile from the BL
lac. After this correction the surrounding nebulosity turned out to be 
well represented by a an elliptical model similarly to the rest of the objects.

\noindent{\bf 1ES 0806+524}: The HST observations of this BL Lac at 
tentative z = 0.136 (Bade et al. 1998) have been discussed by Scarpa
et al (1999a). Noteworthy is an arc--like structure $\sim$2$''$ S of
the nucleus.  Using the NOT data we derive for the host galaxy M(host)
= --24.2 to be compared with M(host) = --23.5 reported by Scarpa et al
(1999a). The difference can be due in part to poor photometry for this
target and to the fact that NOT image is much deeper ($\mu_R$ = 26.5)
than HST image ($\mu_R$ = 24.5) and allow us to detect the host up to
15 arcsec from the center.

\noindent{\bf MS 0922.9+7459}: WSY96 derived for the marginally resolved 
host M(R) = --22.6, while we clearly resolve the elliptical host with M(R) 
= --24.7 and R(e) = 9.7 kpc.  The large difference is likely attributable to 
the difference of seeing (0.7 versus 1.3 arcsec) and therefore the ability 
to distinguish starlight from nuclear source. This high redshift (z = 
0.638) source, which lies behind the rich cluster of galaxies Abell 786 (z 
= 0.124) and may itself be located in a moderately rich cluster (Wurtz et 
al. 1997), was also resolved by HST (U99b), with M(R) = --24.6, in good 
agreement with our result.

\noindent{\bf 1ES 1011+496}: Our derived magnitude for the elliptical 
host galaxy, M(R) = --23.7, is in good agreement with that derived by the 
HST (U99b), M(R) = --23.6. The redshift of this object is uncertain, 
being based on the  possible membership of the BL Lac to the cluster 
Abell 950 at z=0.20 (Wisniewski et al. 1986). Some galaxies are 
indeed detected in the field of view. 

\noindent{\bf 1ES 1028+511}: Our derived magnitude for the elliptical 
host galaxy, M(R) = --23.2, is significantly fainter than that derived by the 
HST (U99b), M(R) = --24.1. 
Again we believe the difference is attributable to different surface brightness limits of the images.
 Note that a 
reliable redshift of z = 0.361 based on CaII H\&K absorption lines 
was recently reported by Polomski et al. (1997), which is considerably 
larger than the z = 0.239 previously used for this target.

\noindent{\bf 1ES 1218+304}: This source has a recently determined redshift 
of z = 0.182 (Bade et al. 1998). WSY96 derived for the host m(R) = 16.6, 
while we derive m(R) = 17.6. HST derived for the 
absolute host magnitude M(R) = --23.6, in good agreement with our result, 
M(R) = --23.8. 

\noindent{\bf MS 1221.8+2452}: U99b derived for the host M(R) = --22.5 
and R$_e$ = 4.0 kpc, WSY96 derived M(R) = --22.7 and R$_e$ = 2.6 kpc, 
and Jannuzi et al (1997) derived M(R) = --22.8 and R$_e$ = 3.2 kpc. We 
derive in this study M(R) = --22.7 and R$_e$ = 2.9 kpc, in good agreement 
with the previous studies.

\noindent{\bf MS 1229.2+6430}: WSY96 derived for the host M(R) = --24.1 
and R(e) = 10.6 kpc, and HST (U99b) derived M(R) = --24.1 while we 
derive M(R) = --24.3 and R(e) = 11.0 kpc, in good agreement with the 
previous determinations. The elliptical host looks quite symmetric 
despite the presence of a companion galaxy located 3.4 arcsec SW.

\noindent{\bf 1ES 1255+244}: Heidt et al (1999) derived for the host M(R) 
= --23.2 and R(e) = 7.2 kpc, and HST (U99b) derived M(R) = --23.3, while 
we derive M(R) = --23.4 and R(e) = 6.4 kpc, in good agreement with 
the previous determinations. This BL Lac seems to be embedded in a 
small cluster of galaxies (Heidt et al. 1999). 

\noindent{\bf MS 1407.9+5954}: U99b using HST derived for the host M(R) 
= --24.8 and R$_e$ = 11.1 kpc, WSY96 derived M(R) = --24.3 and R$_e$ = 9.7 
kpc, while Jannuzi et al (1997) derived M(R) = --24.0 and R$_e$ = 12.2 kpc. 
We derive in this study M(R) = --23.9 and R$_e$ = 8.0 kpc, in good 
agreement with the previous studies. This BL Lac is the BCG in a 
moderately rich cluster (Wurtz et al. 1993, 1997).

\noindent{\bf MS 1443.5+6349}: WSY96 derived for the host which they 
classified as a probable disk, M(R) = --23.6 and R$_e$ = 11.3 kpc. We 
clearly classify the host as an elliptical, with M(R) = --23.9 and R$_e$ 
= 17.2 kpc. This source is surrounded by close companions. 

\noindent{\bf MS 1458.8+2249}: WSY96 derived for the host M(R) = --23.6 
and R$_e$ = 7.9 kpc, and HST (U99b) derived M(R) = --23.7, while we 
derive M(R) = --24.3 and R$_e$ = 2.3 kpc, somewhat brighter host. The 
model fits are hampered by the presence of bright nearby stars.

\noindent{\bf MS 1757.7+7034}: WSY96 derived for the host M(R) = --23.3 
and R$_e$ = 6.1 kpc, and HST (U99b) derived M(R) = --23.6, while we 
derive M(R) = --23.4 and R$_e$ = 2.2 kpc, in good agreement with the 
previous studies.

\noindent{\bf 1ES 1853+671}: Heidt et al (1999) derived for the host M(R) 
= --22.9 and R$_e$ = 9.4 kpc, and HST (U99b) derived M(R) = --23.2, while 
we derive M(R) = --22.7 and R$_e$ = 11.6 kpc, in reasonably good agreement. 
This BL Lac belongs to a very poor group of galaxies, however, it has a 
close companion 2$''$ to the NW (Heidt et al. 1999). 

\noindent{\bf 1ES 1959+650}: The HST observations of this BL Lac at a 
tentative redshift z = 0.048 have been discussed by Scarpa et al (1999a). It 
is hosted by a gas--rich elliptical galaxy with a prominent dust lane. 
They are able to clearly  resolve the host galaxy whose luminosity is 
 M(host) = --22.5. This BL Lac was 
also studied by Heidt et al (1999), who derived for the host galaxy m(host) 
= 14.8, M(host) = --23.0 and R$_e$ = 12.5 kpc. We derive for the host 
galaxy m(host) = 16.1, M(host) = --22.7 and R$_e$ = 5.3 kpc. All 
these determinations are in good agreement with each other, except for 
the scale length. Note that the absolute host luminosity is in the faintest 
end of the distribution for the XBLs, suggesting that its distance could 
be larger than z = 0.048.

\noindent{\bf 1ES 2037+521}: Heidt et al (1999) derive for the host M(R) 
= --23.2 and R$_e$ = 12.3 kpc, while we derive M(R) = --23.7 and R$_e$ = 
7.8 kpc, in reasonable agreement. HST (U99b) derived m(R) = 16.1, in 
good agreement with the value found here, m(R) = 16.3. A small 
companion galaxy is visible 0.6 arcsec away from the nucleus, apparently 
well inside the host. 

\noindent{\bf MS 2143.4+0704}: Kotilainen et al. (1998b) derived for the 
host M(H) = --25.9 and R$_e$ = 5.5 kpc, while we find in this study M(R) 
= --23.6 and R$_e$ = 10.9 kpc. The scale length is in reasonable 
agreement. The color of the host is $R$--$H$ = 2.3, in agreement with 
the average found for low redshift BL Lacs, $R$--$H$ = 
2.2$\pm$0.5 (Kotilainen et al. 1998b). 
Urry et al. (1999a) and U99b derived for the host M(R) = --23.7, M(I) 
= --24.1 and R$_e$ = 8.8 kpc, WSY96 derived M(R) = --23.8 and R$_e$ = 11.6 
kpc, while Jannuzi et al (1997) derived M(R) = --23.3 and R$_e$ = 9,0 kpc, 
in good agreement with our result, M(R) = --23.6.

\noindent{\bf 1ES 2326+174}: Heidt et al (1999) derive for the host M(R) 
= --23.4 and R$_e$ = 8.5 kpc, and HST (U99b) derived M(R) = --23.7,
while we derive M(R) = --24.2 and R$_e$ = 7.5 kpc, slightly brighter
host. Part of the 0.8 mag difference with Heidt et al is probably due
to different extinction and K-correction applied since the difference
of observed galaxy mag is just 0.5 mag. Three
faint galaxies, possibly companions, are superimposed onto the outer
parts of the host at 3--6$''$ distance from the nucleus (Heidt et
al. 1999).

\begin{acknowledgements}
 
This work was partly supported by the Italian Ministry for University and 
Research (MURST) under grant Cofin98-02-32. 
This research has made use of the NASA/IPAC Extragalactic Database (NED) 
which is operated by the Jet Propulsion Laboratory, California Institute of 
Technology, under contract with the National Aeronautics and Space 
Administration. 

\end{acknowledgements}

\noindent{\bf References}

\noindent Abraham,R.G., McHardy,I.M., Crawford,C.S. 1991, MNRAS 252, 482\\
\noindent Antonucci,R. 1993, ARA\&A 31, 473\\
\noindent Antonucci, R.  Ulvestad, J.S. 1985, Ap.J. 294,158\\
\noindent Bade,N., Beckmann,V., Douglas,N.G. \ea 1998, A\&A 334, 459\\
\noindent Bahcall,J.N., Kirhakos,S., Saxe,D.H., Schneider,D.P. 1997, ApJ 479, 642\\
\noindent Blandford,R.D., Rees,M.J. 1978. In  Wolfe A.N. (eds). Pittsburgh 
Conference on BL Lac Objects. University of Pittsburgh Press, p. 328\\
\noindent Bressan,A., Chiosi,C., Fagotto,F. 1994, ApJS 94, 63\\
\noindent Browne, I. W. A. 1983 MNRAS, 204, 23\\
\noindent Browne,I.W.A., Marcha,M.J.M. 1993, MNRAS 261, 795\\
\noindent Cassaro, P., Stanghellini, C., Bondi, M., Dallacasa, D. Della Ceca, R. and Zappala, R.A. 1999, A\&A, in press\\
\noindent Capaccioli,M., Caon,N., D'Onofrio,M. 1992 MNRAS 259, 323\\
\noindent Celotti,A., Maraschi,L., Ghisellini,G., Caccianiga,A., Maccacaro,T. 1993, ApJ 416, 118\\
\noindent Coleman,G.D., Wu,C.C., Weedman,D.W. 1980, ApJS 43, 393\\
\noindent Djorgovski,S., Davis,M. 1987, ApJ 313, 59\\
\noindent Ellingson,E., Yee,H.K.C., Green,R.F. 1991, ApJ 371, 49\\
\noindent Elvis,M., Plummer,D., Schachter,J., Fabbiano,G. 1992, ApJS 80, 257 (E92)\\
\noindent Falomo, R., Pesce, J. E.,  Treves, A. 1993, ApJ 411, L63\\
\noindent Falomo,R., Pesce,J.E., Treves,A. 1995, ApJ 438, L9\\
\noindent Falomo,R. 1996, MNRAS, 283, 241\\
\noindent Falomo,R., Kotilainen,J., Pursimo,T., et al. 1997, A\&A 321, 374\\
\noindent Fanaroff,B., Riley,J.M. 1974, MNRAS 167, 31P\\
\noindent Fukugita,M., Shimasaku,K., Ichikawa,T. 1995, PASP 107, 945\\
\noindent Gioia,I.M., Maccacaro,T., Schild,R.E., et al. 1990, ApJS 72, 567\\
\noindent Govoni, F., Falomo, R, Fasano, G. and Scarpa, R. 1999, A\&A in press\\
\noindent Heckman,T.M. 1990, Paired and Interacting Galaxies (ed. J.W.Sulentic, W.C.Keel, C.M.Telesco), p. 359, NASA Conf. Publ. 3098\\
\noindent Heidt,J., Nilsson,K., Fried, J.W., Takalo,L.O.,  Sillanpaa,A. 1999, A\&A in press\\
\noindent Heidt,J., Nilsson,K., Sillanpaa,A., Takalo,L.O., Pursimo,T. 1999, A\&A 341, 683\\
\noindent Hoessel,J.G., Gunn,J.E., Thuan,T.X. 1980, ApJ 241, 486\\
\noindent Hutchings,J.B., Neff,S.G. 1992, AJ 104, 1\\
\noindent Hutchings,J.B., Holtzman,J., Sparks,W.B. et al. 1994, ApJ 429, L1\\
\noindent Jannuzi,B.T., Yanny,B., Impey,C. 1997, ApJ 491, 146\\
\noindent Kollgaard,R.I., Wardle,J.F.C., Roberts,D.H., Gabuzda,D.C. 1992, AJ 104, 1687\\
\noindent Kotilainen,J.K., Falomo,R., Scarpa,R. 1998a, A\&A 332, 503\\
\noindent Kotilainen,J.K., Falomo,R., Scarpa,R. 1998b, A\&A 336, 479\\
\noindent Lamer, G., Newsam, A.M., McHardy,I.M. 1999, MNRAS, in press\\
\noindent Lamer, G., McHardy,I.M. Newsam, A.M.  1999. In L.O.Takalo, A.Sillanp\"a\"a (eds) BL Lac Phenomenon. ASP Conf. Ser. 159, 381\\
\noindent Landolt,A.U. 1992, AJ 104, 340\\
\noindent Lehnert,M.D., Heckman,T.M., Chambers,K.C., Miley,G.K. 1992, ApJ 393, 68\\
\noindent Maccacaro,T., Wolter,A., McLean,B., et al. 1994, Astroph. Lett. Comm. 29, 267\\
\noindent Magorrian,J., Tremaine,S., Richstone,D., et al. 1998, AJ 115, 2285\\
\noindent McHardy,I.M., Abraham,R.G., Crawford,C.S. et al. 1991, MNRAS 249, 742\\
\noindent McHardy,I.M., Luppino,G.A., George,I.M., Abraham,R.J., Cooke,B.A. 1992, MNRAS 256, 655\\
\noindent McLure,R.J., Dunlop,J.S., Kukula,M.J. et al. 1999, ApJ in press\\
\noindent Mobasher,B., Sharples,R.M., Ellis,R.S. 1993, MNRAS 263, 560\\
\noindent Morris,S.L., Stocke,J.T., Gioia,I.M., et al. 1991, ApJ 380, 49 (M91)\\
\noindent Murphy D.W., Browne W.A., PerleyR.A. 1993, MNRAS 264, 298\\
\noindent Ostriker,J.P., Vietri,M. 1990, Nat 344, 45\\
\noindent Owen,F.N., Laing,R.A. 1989, MNRAS 238, 357\\
\noindent Owen,F.N., Ledlow,M.J., Keel,W.C. 1996, AJ 111, 53\\
\noindent Padovani,P. 1992, MNRAS 257, 404\\
\noindent Padovani,P., Urry,C.M. 1990, ApJ 356, 75\\
\noindent Padovani,P., Giommi,P. 1995, MNRAS 277, 1477\\
\noindent Perlman,E.S., Stocke,J.T. 1993, ApJ 406, 430\\
\noindent Perlman,E.S., Stocke,J.T., Schachter,J.F., et al. 1996, ApJS 104, 251 (P96)\\
\noindent Pesce,J.E., Falomo,R., Treves,A. 1995 AJ 110, 1554\\
\noindent Polomski,E., Vennes,S., Thorstensen,J.P., Mathioudakis,M., Falco,E.E. 1997, ApJ 486, 179\\
\noindent R\"onnback,J., van Groningen,E., Wanders,I., \"Orndahl,E. 1996, MNRAS 283, 282\\
\noindent Scarpa,R., Urry,C.M., Falomo,R., et al. 1999a, ApJ 521, 134\\
\noindent Scarpa,R., et al. 1999b, ApJS in press\\
\noindent Shull,J.M., Van Steenberg,M.E. 1985, ApJ 294, 599\\
\noindent Stanghellini, C., Dallacasa, D., Bondi, M., and Della Ceca,R. 1997, A\&A 325,911\\
\noindent Stark,A.A., Gammie,C.F., Wilson,R.W., et al. 1992, ApJS 79, 77\\
\noindent Stocke,J.T., Morris,S.L., Gioia,I.M., et al. 1991, ApJS 76, 813\\
\noindent Stocke, J.T., Wurtz, R.E., Wang, Q.D.,  Elston, R.,  Januzzi, B.T. 1992, Ap.J, 400, L17\\
\noindent Stocke,J.T., Wurtz,R.E., Perlman,E.S. 1995, ApJ 454, 55\\
\noindent Thuan,T.X., Romanishin,W. 1981, ApJ 248, 439\\
\noindent Urry,C.M., Padovani,P. 1995, PASP 107, 803\\
\noindent Urry,C.M., Falomo,R., Scarpa,R. et al. 1999a, ApJ 512, 88\\
\noindent Urry et al 1999b, ApJ, in press (U99b)\\
\noindent Wisniewski,W.Z., Sitko,M.L. Sitko,A.K. 1986 MNRAS 219, 299\\
\noindent Wardle, J.F.C., Moore, R.L., and Angel, J.R.P. 1984, Ap.J  279, 93\\
\noindent Wurtz,R., Ellingson,E., Stocke,J.T., Yee,H.K.C. 1993, AJ 106, 869\\
\noindent Wurtz,R., Stocke,J.T., Yee,H.K.C. 1996, ApJS 103, 109 (WSY96)\\
\noindent Wurtz,R., Stocke,J.T., Ellingson,E., Yee,H.K.C. 1997, ApJ 480, 547

\end{document}